\documentclass[aps,prl,reprint,superscriptaddress]{revtex4-2}

\usepackage{graphicx} 
\usepackage{amsmath}  
\usepackage{hyperref} 
\usepackage{siunitx}  

\begin{document}

\title{Extending \texttt{SLUSCHI} for Automated Diffusion Calculations}

\author{Qi-Jun Hong}
\email[]{qhong@alumni.caltech.edu}
\affiliation{Materials Science and Engineering, School for Engineering of Matter, Transport, and Energy, Arizona State University, Tempe, Arizona 85285, USA}

\author{Qing Chen}
\affiliation{Thermo-Calc Software AB, Råsundavägen 18, SE-169 67, Solna, Sweden}
\affiliation{KTH Royal Institute of Technology, SE-100 44, Stockholm, Sweden}

\author{Ligen Wang}
\affiliation{Materials Science and Engineering, School for Engineering of Matter, Transport, and Energy, Arizona State University, Tempe, Arizona 85285, USA}

\author{Dallin Fisher}
\affiliation{Materials Science and Engineering, School for Engineering of Matter, Transport, and Energy, Arizona State University, Tempe, Arizona 85285, USA}

\author{Audrey CampBell}
\affiliation{Materials Science and Engineering, School for Engineering of Matter, Transport, and Energy, Arizona State University, Tempe, Arizona 85285, USA}

\author{Si-Da Xue}
\affiliation{Materials Science and Engineering, School for Engineering of Matter, Transport, and Energy, Arizona State University, Tempe, Arizona 85285, USA}

\author{Linqin Mu}
\affiliation{Materials Science and Engineering, School for Engineering of Matter, Transport, and Energy, Arizona State University, Tempe, Arizona 85285, USA}

\author{Noemi Leick}
\affiliation{National Renewable Energy Laboratory, Golden, CO 80401, USA}

\author{Seetharaman Sridhar}
\affiliation{Materials Science and Engineering, School for Engineering of Matter, Transport, and Energy, Arizona State University, Tempe, Arizona 85285, USA}

\date{\today}

\begin{abstract}
    We present an extension of the \texttt{SLUSCHI} package (Solid and Liquid in Ultra Small Coexistence with Hovering Interfaces) to enable automated diffusion calculations from first-principles molecular dynamics. While the original \texttt{SLUSCHI} workflow was designed for melting temperature estimation via solid–liquid coexistence, we adapt its input and output handling to isolate the volume search stage and generate one production trajectory suitable for diffusion analysis. Post-processing tools parse \texttt{VASP} outputs, compute mean-square displacements (MSD), and extract tracer diffusivities using the Einstein relation with robust error estimates through block averaging. Diagnostic plots, including MSD curves, running slopes, and velocity autocorrelations, are produced automatically to help identify diffusive regimes. The method has been validated through representative case studies: self- and inter-diffusion in Al–Cu liquid alloys, sublattice melting in Li$_7$La$_3$Zr$_2$O$_{12}$ and Er$_2$O$_3$, interstitial oxygen transport in bcc and fcc Fe, and oxygen diffusivity in Fe–O liquids with variable Si and Al contents. Viscosity and diffusivity are linked through the Stokes–Einstein relation, with composition dependence assessed via simple linear mixing. This capability broadens \texttt{SLUSCHI} from melting-point predictions to transport property evaluation, enabling high-throughput, fully first-principles datasets of diffusion coefficients and viscosities across metals and oxides.
\end{abstract}
\maketitle

\section*{Introduction}
First-principles density functional theory (DFT) has become a standard framework for quantifying atomic diffusion in condensed phases. Two complementary families of approaches are widely used \cite{Sandberg2002PRL, Mantina2008PRL, Mantina2009Acta, Blochl1993PRL, Milman1993PRL}. (i) Static, rare-event techniques \cite{Sandberg2002PRL, Mantina2008PRL, Mantina2009Acta} based on transition-state theory (TST) evaluate migration barriers and prefactors along atomistic pathways, typically via the nudged elastic band (NEB) method and its climbing-image and improved-tangent variants \cite{Henkelman2000CI,Henkelman2000NEB}. These tools are broadly available in mainstream plane-wave codes. (ii) \textit{Ab initio} molecular dynamics (AIMD) \cite{Blochl1993PRL, Milman1993PRL} directly samples finite-temperature transport in liquids and disordered solids, enabling the computing of diffusion coefficients and structural correlations. For long-time kinetics that lie beyond straightforward AIMD, rare-event frameworks leveraging on-the-fly saddle searches (e.g., the dimer method) coupled with kinetic Monte Carlo can bridge to microseconds and beyond \cite{Henkelman1999Dimer}.

Although several workflow managers and analysis packages exist that partially automate DFT–MD and diffusion analysis, they typically require stitching together separate tools or custom scripting. For example, the Materials Simulation Toolkit (MAST) provides high-level, defect- and diffusion-focused workflows for \texttt{VASP}-based NEB and MD calculations \cite{Mayeshiba2017MAST}, and the Atomate framework (with MPmorph AIMD workflows) supplies well-tested, reusable workflow templates and AIMD-oriented tooling built on the Materials Project stack \cite{Mathew2017Atomate,MPmorph2023}. The Atomic Simulation Environment (ASE) offers flexible Python steering of MD and calculator backends, making it a common choice for building custom automated MD→MSD pipelines \cite{Larsen2017ASE}, while commercial packages such as QuantumATK provide an integrated GUI/CLI environment with built-in AIMD and diffusion analysis utilities for users preferring an out-of-the-box solution \cite{Smidstrup2020QuantumATK}. \texttt{VASPKIT} can directly parse VASP MD outputs and compute MSD and related transport quantities, thereby streamlining the MD → diffusivity analysis pipeline \cite{Geng2025VASPKIT,Wang2021VASPKIT}. The present \texttt{"SLUSCHI}–Diffusion" module complements these options by combining the thermal-expansion/volume-search automation with turnkey trajectory parsing and automated MSD → diffusivity extraction in a single pipeline.

Within this landscape, \texttt{SLUSCHI} was originally developed for first-principles {melting-temperature} calculations using small-cell solid–liquid coexistence \cite{Hong2013SmallCoex, Hong2016}. In this work we re-purpose and extend the same DFT–MD infrastructure specifically for {diffusion} in regimes where atomic mobility is intrinsically fast—namely (i) liquids and (ii) crystalline solids exhibiting sublattice melting. Rather than inferring the melting point ($T_\mathrm{m}$) from coexistence, we run \textit{NVT}/\textit{NPT} trajectories directly in the target phase(s), parse unwrapped species-resolved trajectories, and compute transport metrics (e.g., MSD and self-diffusion coefficients) with an automated pipeline. This “\texttt{SLUSCHI}–Diffusion” workflow preserves \texttt{SLUSCHI}’s automation (input generation, run orchestration, and post-processing) while focusing squarely on diffusion in high-mobility phases.

\section{Methods}

\begin{figure}
    \centering
    \includegraphics[width=0.49\textwidth]{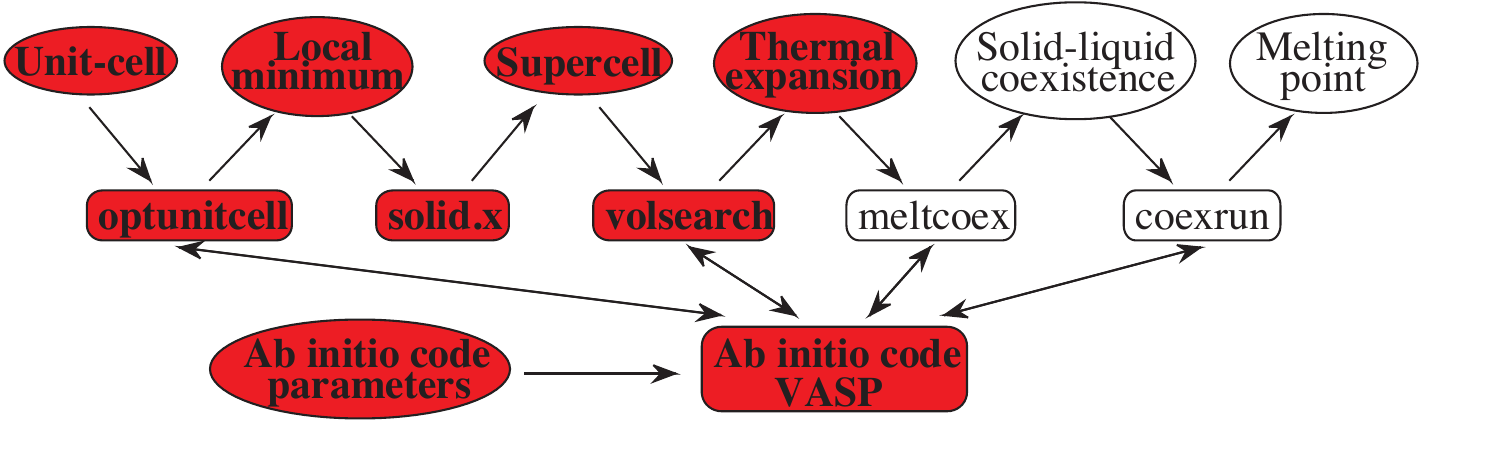} 
    \caption{%
Workflow of the \texttt{SLUSCHI} framework. 
The original modules (\texttt{optunitcell}, \texttt{volsearch}, \texttt{meltcoex}, and \texttt{coexrun}) are shown schematically to highlight their sequence from unit-cell optimization and supercell construction through thermal expansion, solid–liquid coexistence, and melting-point fitting. 
For diffusion calculations, only the first two stages (unit-cell optimization and volume search/thermal expansion) are executed, generating one long \textit{ab initio} molecular dynamics trajectory of equilibrated liquid or partially disordered configurations for subsequent automated diffusion analysis. 
This modular design allows the same infrastructure to perform both melting and high-temperature transport calculations within a unified, fully automated DFT–MD workflow.}
    \label{Figures:diagram}
\end{figure}

\subsection{Calculation of diffusion coefficient}

Our diffusion calculations are based on AIMD performed with \texttt{VASP} and orchestrated through the \texttt{SLUSCHI} framework. \texttt{SLUSCHI} automates the preparation of liquid or coexistence structures, equilibration at the target thermodynamic state, and long production trajectories, making it possible to gather reliable atomic displacement data without manual intervention. For both solid and liquid systems, the \textit{NPT} ensemble is employed with periodic cell-shape relaxations to ensure proper equilibration. Simulation lengths are chosen to capture at least tens of picoseconds of diffusive motion, ensuring convergence of MSD rate.

Diffusion coefficients are obtained from the atomic trajectories using the Einstein--Smoluchowski relation of Brownian motion theory. Specifically, the MSD of each species is computed as a function of time, and the slope of the MSD in the linear (diffusive) regime is related to the self-diffusion coefficient \(D^{\text{self}}_\alpha\) of species alpha via
\begin{equation}
D_\alpha = \frac{1}{2d}\,\frac{d}{dt}\left\langle \big|\mathbf{r}_i(t+t_0)-\mathbf{r}_i(t_0)\big|^2 \right\rangle_{t_0}, \quad d=3.
\end{equation}
This procedure is implemented in \texttt{SLUSCHI}’s diffusion module, which also performs block averaging \cite{Flyvbjerg1989BlockAverage, FrenkelSmit2002Book} to quantify statistical error bars. The method requires no empirical input parameters, relying solely on the atomic motions predicted by DFT-based AIMD, and therefore provides a robust and transferable way to compute diffusivities across different compositions and phases~\cite{Hong2016}.

The value of this methodology is that it captures quantitative trends in diffusion under specific conditions that are often inaccessible to experiment. Experimental diffusivity measurements are typically limited to dilute concentrations and ambient conditions, whereas AIMD with \texttt{SLUSCHI} allows us to probe non-dilute alloys, high temperatures, and complex liquid states. Such simulations thus provide a rigorous and quantitative benchmark for diffusion, complementing experimental data and clarifying how structural phase transitions and composition affect atomic mobility.

The diffusion capability provides automated post-processing of \texttt{VASP} MD trajectories to compute:
(1) self-diffusion coefficients \(D^{\text{self}}_\alpha\) for each atomic species via the Einstein relation,
(2) species-resolved MSD and velocity‐autocorrelation diagnostics,
(3) uncertainty quantification from block averaging and windowed linear fits in the diffusive regime.
The module plugs into \texttt{SLUSCHI}’s job orchestration so our users can launch, monitor, and analyze MD runs without manual parsing.

\begin{table*}
\caption{\label{jobin_example}Example \texttt{SLUSCHI} \texttt{job.in} for running diffusion calculations of liquid FeO at 2063 K.}
\begin{verbatim}
temp = 2063.0 # Target temperature in K 
press = 0.0 # Pressure in kbar (0 = ambient)

# Barostat settings:
navg = 3 # average pressure computed over (80 × navg) MD steps
factor = 10000.0 # barostat "stiffness"; large = sluggish response to stress

# Thermal expansion control:
thmexp_only = 1 # Only do thermal expansion stage (1 = yes, 0 = run full workflow)
thmexp_cnvg = 0.01 # Error tolerance for convergence of pressure
thmexp_max = 200 # Maximum (80 × thmexp_max) steps for thermal expansion
thmexp_min = 1 # Minimum steps
thmexp_liq = 1 # Conduct volsearch using a liquid reference (1 = yes)
thmexp_l_t = 4000 # Initial temperature to melt solid and ensure a liquid phase

vaspcmd = sbatch # Job command for VASP submission (e.g., sbatch, qsub)

kmesh = -1 # kmesh: -1 triggers the special (¼,¼,¼) low-cost k-point option
radius = 12.0 # Cell construction: target minimum image distance (Å)

# Advanced settings:
# tgt_nelm and adj_bmix refer to POTIM/BMIX adjustment and hybrid step control
tgt_nelm = 6
adj_bmix = 1
# adj_potim = 0 disables POTIM auto-tuning (i.e., time step adjustments)
adj_potim = 0
\end{verbatim}
\end{table*}

\begin{table*}
\centering
\caption{Example of the standard output from the \texttt{SLUSCHI}–Diffusion module showing parsed results for AlCu liquid at 2000 K. The output is printed automatically to \texttt{diffusion.out}.}
\label{tab:diffusion_output_raw}
\begin{verbatim}
natoms:  184
Element # 1
Total MD length: 		43197.00 fs
Diffusion coefficient is: 	7.81e-04 Ang^2/fs or 7.81e-05 cm^2/s
R2 of the linear fitting is: 	0.99
The uncertainty of the slope is: 	8.36e-08
Element # 2
Total MD length: 		43197.00 fs
Diffusion coefficient is: 	9.52e-04 Ang^2/fs or 9.52e-05 cm^2/s
R2 of the linear fitting is: 	0.98
The uncertainty of the slope is: 	1.40e-07
\end{verbatim}
\end{table*}

\subsection{Requirements, Workflow, and Outputs}

To use the diffusion module of \texttt{SLUSCHI} you need: a \texttt{VASP} installation with PAW potentials, \texttt{SLUSCHI} configured with job submission on your cluster (e.g.\ via \texttt{sbatch} or \texttt{qsub}), and a working Python 3 environment for analysis. You also should have control over key parameters in \texttt{job.in}, such as \texttt{radius} (supercell size), and \texttt{kmesh}, so that MD runs are well-behaved and yield meaningful diffusivities and viscosities.  

Typical MD settings are:
\begin{itemize}
  \setlength\itemsep{0em}      
  \setlength\parindent{0pt}    
  \setlength\leftskip{0pt}     
  \setlength\itemindent{0pt}   
  \setlength\labelsep{0.5em}   
  \item Ensemble: \texttt{SLUSCHI} uses canonical MD (\textit{NVT}) for 80 ionic steps; barostat (\textit{NPT}-like) accomplished by adjusting cell shape and volume every 80 steps using average pressure vs target pressure. 
  \item Timestep: provided in \texttt{INCAR} \texttt{POTIM}; optionally auto-adjusted by \texttt{SLUSCHI} if \texttt{adj\_potim} is enabled.
  \item Supercell size is controlled by the \texttt{radius} tag in \texttt{job.in}; sufficient size is needed to avoid finite-size effects.
  \item K-mesh: specified via the \texttt{kmesh} tag in \texttt{job.in}; when \texttt{kmesh = -1}, \texttt{SLUSCHI} uses a special (¼,¼,¼) low-cost k-point scheme.
\end{itemize}

The following describes the streamlined workflow for performing diffusion calculations using \texttt{SLUSCHI}. Unlike the full melting-temperature sequence, this procedure uses only the volume search stage, and concludes before melting or coexistence simulations. The steps below assume you have prepared required input files and job settings (e.g.\ temperature, pressure, supercell size) in \texttt{job.in}, and that you are ready to run molecular dynamics trajectories for the compositions of interest.

\begin{enumerate}
  \setlength\itemsep{0em}      
  \setlength\parindent{0pt}    
  \setlength\leftskip{0pt}     
  \setlength\itemindent{0pt}   
  \setlength\labelsep{0.5em}   
  \item Prepare standard \texttt{VASP} inputs (\texttt{INCAR}, \texttt{POSCAR}, \texttt{POTCAR}). Leave \texttt{KPOINTS} optional if using auto K-mesh via \texttt{job.in}.
  \item In \texttt{job.in}, set target temperature, pressure, and other control tags (e.g.\ \texttt{radius}, \texttt{adj\_potim}, \texttt{adj\_bmix}, \texttt{tgt\_nelm}) for each composition.
  \item Launch \texttt{SLUSCHI} to generate/execute the MD trajectory in \texttt{Dir\_VolSearch} (volume search stage).
\end{enumerate}

After the MD trajectory completes, invoke \texttt{diffusion.csh} or \texttt{diffusion\_ensemble.csh} to run the diffusion analysis. The automated diffusion module proceeds as follows:

\begin{enumerate}
  \setlength\itemsep{0em}      
  \setlength\parindent{0pt}    
  \setlength\leftskip{0pt}     
  \setlength\itemindent{0pt}   
  \setlength\labelsep{0.5em}   
  \item \textbf{Parsing}: Read \texttt{OUTCAR} to extract per-species unwrapped atomic trajectories and time grid.
  \item \textbf{MSD calculation}: For each species $\alpha$, and N$_{\alpha}$ the number of species, compute
  \[
    \mathrm{MSD}_\alpha(t) = \frac{1}{N_\alpha} \sum_{i \in \alpha} \left\langle \big\lvert \mathbf{r}_i(t_0 + t) - \mathbf{r}_i(t_0) \big\rvert^2 \right\rangle_{t_0}.
  \]
  \item \textbf{Diffusivity (Einstein relation)}: Fit the MSD vs.\ time curve in a user-chosen window $[t_1, t_2]$ within the diffusive regime,
  \[
    D_\alpha = \frac{1}{2d}\,\frac{d\,\mathrm{MSD}_\alpha(t)}{dt}\;\Bigg|_{t\,\in\,[t_1,t_2]}, \quad d=3.
  \]
  \item \textbf{Error bars}: Perform block averaging over multiple overlapping windows; report mean $\pm$ standard error.
  \item \textbf{Diagnostics}: Generate and inspect plots of MSD, running slope, and optionally velocity autocorrelation functions; flag any non-diffusive behavior (e.g.\ sub-diffusion, drift).
\end{enumerate}

\begin{itemize}
  \setlength\itemsep{0em}      
  \setlength\parindent{0pt}    
  \setlength\leftskip{0pt}     
  \setlength\itemindent{0pt}   
  \setlength\labelsep{0.5em}   
  \item Display of species diffusion coefficients on the terminal/screen.
  \item \texttt{diffusion\_coef.png}: plot of MSD vs.\ time with linear fit overlay and chosen fit window.
\end{itemize}

\subsection*{Example job.in with \texttt{SLUSCHI} Comments}

Shown in Table \ref{jobin_example} is a sample \texttt{job.in} for a \texttt{SLUSCHI} run, annotated inline to explain each tag according to the original user guide~\cite{Hong2016}.

The original \texttt{SLUSCHI} workflow consists of four stages (in directories: \texttt{Dir\_OptUnitCell}, \texttt{Dir\_VolSearch}, \texttt{Dir\_Melt}, \texttt{Dir\_CoexRun}). For diffusion calculations, we only use the first two stages—volume search (or thermal expansion, in \texttt{Dir\_VolSearch})—and the program terminates before the melt or coexistence runs, which are intended for melting-temperature determination. MD are conducted under canonical ($NVT$) conditions, but an effective barostat is applied by adjusting the lattice vectors every 80 ionic MD steps based on the average pressure relative to the target pressure. If \texttt{thmexp\_liq} is set to 1, then the first five MD jobs (totaling 400 ionic steps) are launched at a higher temperature (either the value of \texttt{thmexp\_l\_t} or \(2\times\) the target temperature if \texttt{thmexp\_l\_t} is not provided) to ensure the supercell melts.

In \texttt{SLUSCHI}, the tag \texttt{adj\_bmix} determines whether the \texttt{BMIX} mixing parameter in the \texttt{VASP} \texttt{INCAR} is adjusted automatically; \texttt{adj\_potim} controls whether \texttt{POTIM} (the MD time‐step size) is tuned, while \texttt{tgt\_nelm} sets the target number of electronic iterations per MD step when \texttt{POTIM} adjustment is enabled~\cite{Hong2016}. The option \texttt{kmesh = -1} activates a special \((\tfrac{1}{4},\tfrac{1}{4},\tfrac{1}{4})\) k-point scheme, reducing computational cost, yet preserving good accuracy for melting‐related properties~\cite{Hong2016}. All tags are defined in the \texttt{SLUSCHI} user guide for reference if needed \cite{Hong2016}.

\section{Results and Discussion}

\begin{table}[t]
\centering
\caption{Liquid Al–Cu at 1173, 1500, and 2000 K: tracer diffusion coefficients (in units of $10^{-5}$~cm$^2$/s) and viscosity. Composition is $A_xB_{1-x}$ with $A=\mathrm{Al}$ and $B=\mathrm{Cu}$. Uncertainties are one standard deviation.}
\label{tab:alcu_1173_1500_2000K}
\begin{tabular}{lccc}
\hline
\multicolumn{4}{c}{\textbf{1173 K}} \\
Composition & $D_A$ & $D_B$ & $\eta$ (mPa$\cdot$s) \\
\hline
$\mathrm{Al}_{1.00}$                   & $8.60 \pm 0.73$ & ---           & $0.71 \pm 0.06$ \\
$\mathrm{Al}_{0.75}\mathrm{Cu}_{0.25}$ \hspace{.1cm} & \hspace{.1cm} $5.85 \pm 0.31$ \hspace{.1cm} & \hspace{.1cm} $6.00 \pm 0.95$ \hspace{.1cm} & \hspace{.1cm} $1.06 \pm 0.09$ \hspace{.1cm} \\
$\mathrm{Al}_{0.50}\mathrm{Cu}_{0.50}$ & $2.62 \pm 0.64$ & $2.94 \pm 0.12$ & $2.29 \pm 0.32$ \\
$\mathrm{Al}_{0.25}\mathrm{Cu}_{0.75}$ & $1.46 \pm 0.13$ & $1.59 \pm 0.06$ & $4.22 \pm 0.36$ \\
$\mathrm{Cu}_{1.00}$                   & ---              & $1.52 \pm 0.25$ & $4.35 \pm 0.71$ \\
\hline
\multicolumn{4}{c}{\textbf{1500 K}} \\
Composition & $D_A$ & $D_B$ & $\eta$ (mPa$\cdot$s) \\
\hline
$\mathrm{Al}_{1.00}$                   & $14.1 \pm 0.7$ & ---             & $0.56 \pm 0.03$ \\
$\mathrm{Al}_{0.75}\mathrm{Cu}_{0.25}$ & $11.1 \pm 0.4$ & $9.2 \pm 1.7$   & $0.76 \pm 0.06$ \\
$\mathrm{Al}_{0.50}\mathrm{Cu}_{0.50}$ & $6.4 \pm 0.6$  & $5.8 \pm 0.6$   & $1.35 \pm 0.12$ \\
$\mathrm{Al}_{0.25}\mathrm{Cu}_{0.75}$ & $3.9 \pm 0.6$  & $3.6 \pm 0.3$   & $2.26 \pm 0.19$ \\
$\mathrm{Cu}_{1.00}$                   & ---             & $4.0 \pm 0.6$   & $2.09 \pm 0.27$ \\
$\mathrm{Cu}_{1.00}$, LDA              & ---             & $2.6 \pm 0.3$   & $3.22 \pm 0.27$ \\
\hline
\multicolumn{4}{c}{\textbf{2000 K}} \\
Composition & $D_A$ & $D_B$ & $\eta$ (mPa$\cdot$s) \\
\hline
$\mathrm{Al}_{1.00}$                   & $24.5 \pm 2.8$ & ---             & $0.43 \pm 0.05$ \\
$\mathrm{Al}_{0.75}\mathrm{Cu}_{0.25}$ & $17.2 \pm 2.2$ & $20.0 \pm 1.6$  & $0.60 \pm 0.07$ \\
$\mathrm{Al}_{0.50}\mathrm{Cu}_{0.50}$ & $12.4 \pm 1.7$ & $12.3 \pm 0.7$  & $0.88 \pm 0.08$ \\
$\mathrm{Al}_{0.25}\mathrm{Cu}_{0.75}$ & $8.3 \pm 2.4$  & $9.2 \pm 1.1$   & $1.24 \pm 0.20$ \\
$\mathrm{Cu}_{1.00}$                   & ---             & $7.8 \pm 0.6$   & $1.44 \pm 0.12$ \\
\hline
\end{tabular}
\end{table}

\begin{figure}
    \centering
    \includegraphics[width=0.49\textwidth]{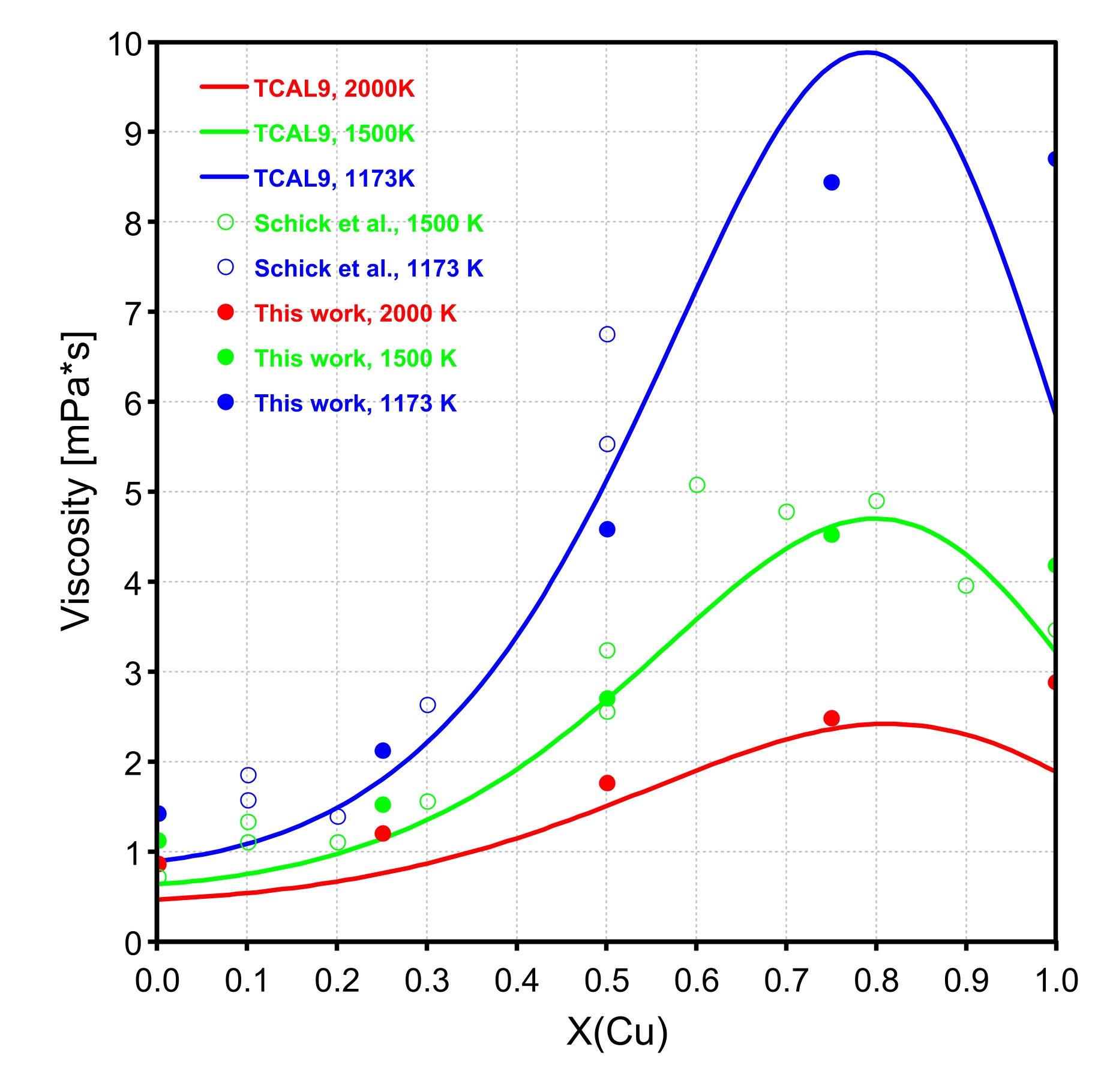} 
    \caption{Viscosity (scaled by a factor of 2 due to PBE underbinding nature) of Al–Cu liquids at various compositions at 1173, 1500, and 2000 K, compared with experimental data \cite{Schick2012} and CALPHAD calculations using the TCAL9 thermodynamic database \cite{TCAL9_2025}.
    }
    \label{fig:vis_comp}
\end{figure}

\subsection{Diffusion and viscosity in liquid Al--Cu system}

In this work, we relate tracer diffusivities to viscosity using the Stokes--Einstein (SE) relation,
\[
D_i \;=\; \frac{k_\mathrm{B}T}{c \,\pi\, \eta_i \, r_i}\,,
\]
with $c{=}6$ for stick boundary conditions, $\eta_i$ the shear viscosity, and $r_i$ the atomic radius for species $i$~\cite{Costigliola2019}. For binary Al--Cu liquids we adopt a pragmatic linear mixing for quantities that enter SE,
\[
\eta(x_\mathrm{Cu}) \approx (1{-}x_\mathrm{Cu})\,\eta_\mathrm{Al} + x_\mathrm{Cu}\,\eta_\mathrm{Cu}, 
\]
so that $D_i(T,x)$ follows directly from measured or computed $\eta(T,x)$ and the chosen $c$ and $r_i$. This linear “rule of mixtures’’ is used here to construct smooth $D$–vs–composition datasets from endpoint (pure-component) inputs and our viscosity table. We note that metallic melts can deviate from ideal mixing and from exact SE scaling (e.g., composition-dependent excess viscosity or clustering near certain $x_\mathrm{Cu}$), but the SE+linear approach provides a transparent, first-order bridge between $\eta$ and $D$ suitable for comparison across compositions and temperatures~\cite{Schick2012,Pan2017,FeCoNi2021}.

Figure~\ref{fig:vis_comp} compares the \textit{ab initio} viscosities of liquid Al–Cu alloys obtained from the Stokes–Einstein relation with the CALPHAD-type assessments of Schick \textit{et al.}~\cite{Schick2012} across a range of compositions and temperatures. The DFT–MD results reproduce both the qualitative and quantitative trends of the CALPHAD model: viscosity increases with Cu-content and exhibits a pronounced maximum near intermediate compositions, reflecting stronger short-range chemical ordering in Cu-rich melts. The magnitude of $\eta$ agrees with CALPHAD predictions in the temperature range, a level consistent with typical uncertainties in both experimental and thermodynamic-model data for metallic liquids. The temperature dependence follows an Arrhenius-like decrease, with activation energies derived from \texttt{SLUSCHI}–Diffusion trajectories lying between 0.25–0.35 eV, close to the experimentally inferred values. These consistencies confirm that the automated DFT–MD workflow not only reproduces absolute transport magnitudes but also captures the correct composition and temperature trends encoded in CALPHAD-based viscosity databases.

It should be noted that all calculated DFT viscosities were multiplied by a scaling factor of two to correct for the systematic underestimation of viscosity inherent to the Perdew–Burke–Ernzerhof (PBE) \cite{Perdew1996PBE} exchange–correlation functional. The semi-local PBE approximation is known to underbind metallic and covalent liquids, resulting in slightly overexpanded interatomic distances and reduced short-range order, which in turn yield artificially low viscosities and excessively high diffusivities in AIMD simulations. Applying this empirical factor brings the absolute magnitudes of the simulated viscosities into quantitative agreement with experimental and CALPHAD-based data, while retaining the correct composition and temperature dependencies predicted by first-principles theory. In addition, one test case (Table \ref{tab:alcu_1173_1500_2000K}) computed using the local-density approximation (LDA) for pure Cu–Al yields a viscosity that agrees very well with experiment, confirming that the magnitude of the PBE scaling correction is physically consistent.

\begin{figure*}
    \centering
    \includegraphics[width=0.99\textwidth]{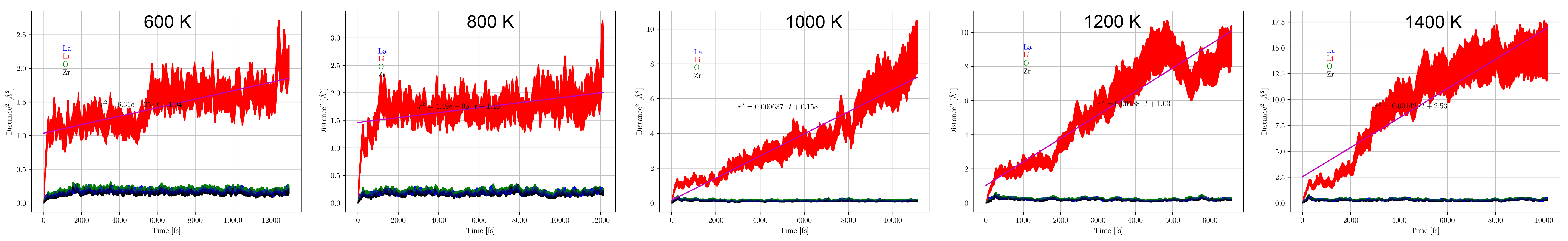} 
    \caption{Diffusion coefficient of lithium in LLZO from 600 to 1400 K. Sublattice melting occurs between 800 and 1000 K.
    }
    \label{Figures:LLZO_diff}
\end{figure*} 

\begin{figure}
    \centering
    \includegraphics[width=0.49\textwidth]{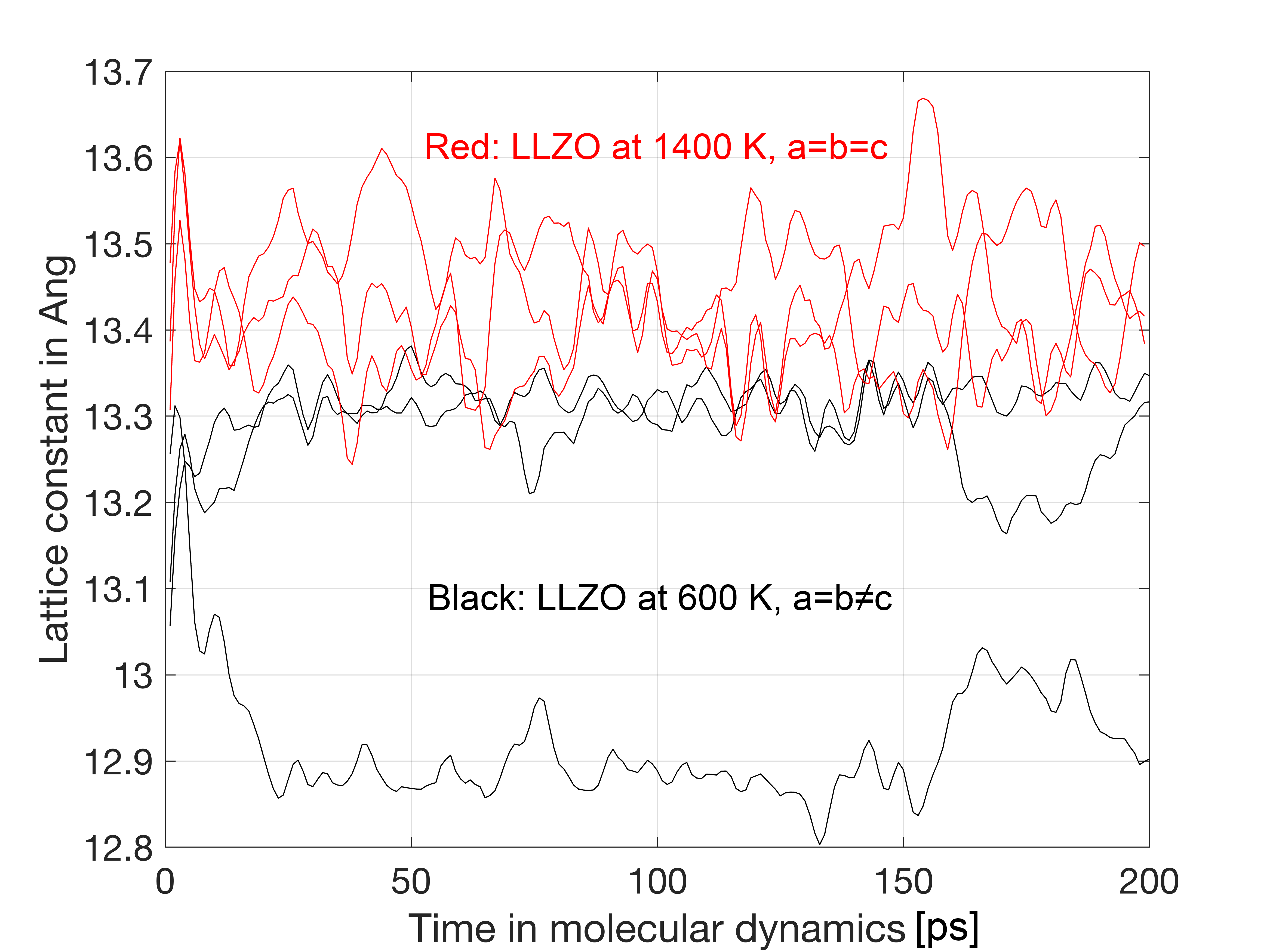} 
    \caption{Lattice constants $a$, $b$, and $c$ in LLZO at 600 K (balck trace) and 1400 K (red trace).
    }
    \label{Figures:LLZO_latt}
\end{figure} 

\begin{figure}
    \centering
    \includegraphics[width=0.49\textwidth]{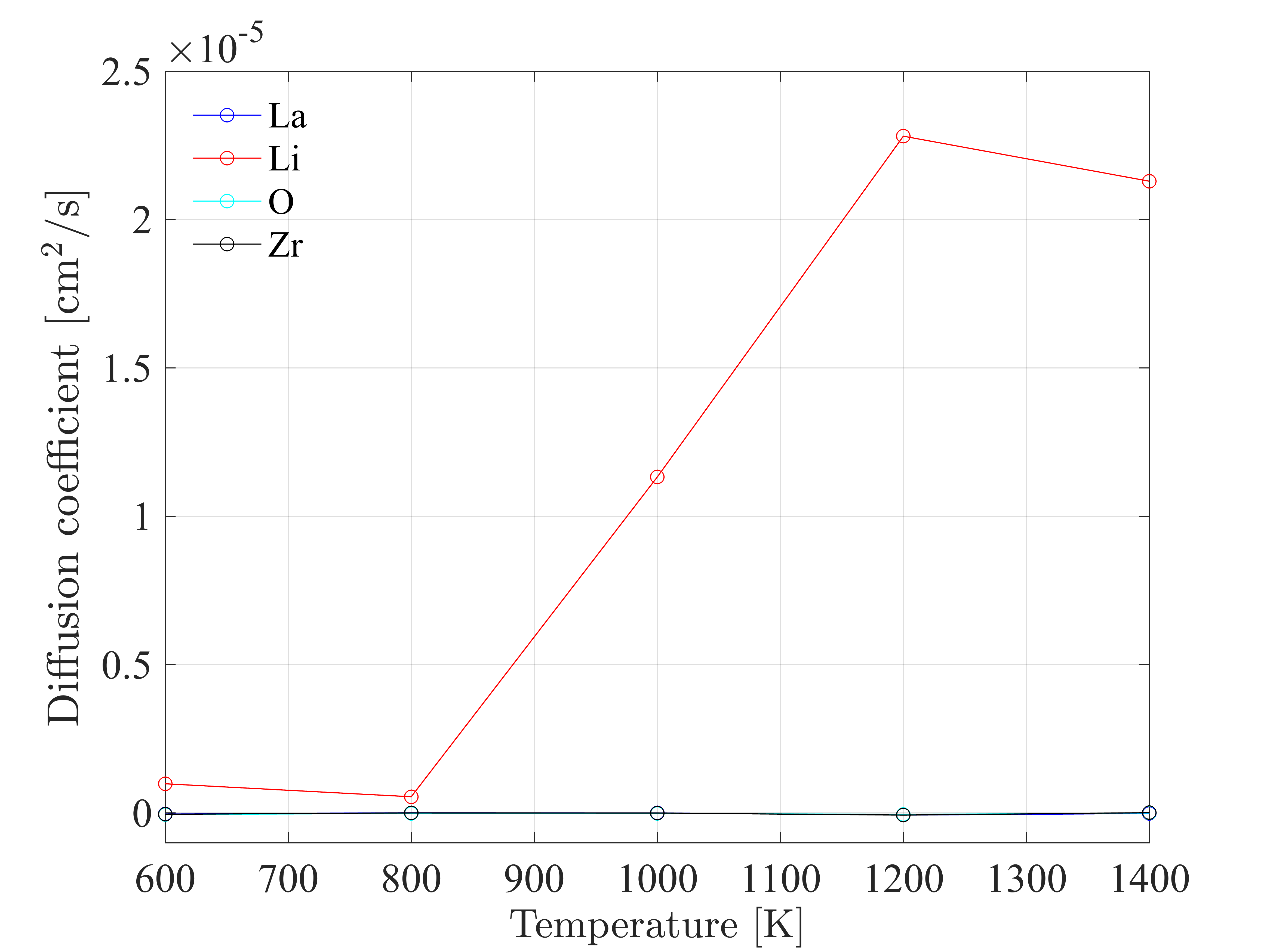} 
    \caption{Diffusion coefficients in LLZO from 600 to 1400 K.
    }
    \label{Figures:LLZO_diff}
\end{figure} 


\subsection*{Sublattice melting in LLZO and H-phase Er$_2$O$_3$}

\subsubsection{Lithium Sublattice Melting and \\the Tetragonal-to-Cubic Transition in LLZO}

Our DFT-MD simulations clearly demonstrate that the tetragonal-to-cubic (t--c) phase transition in Li$_7$La$_3$Zr$_2$O$_{12}$ (LLZO) is driven by the progressive melting of the lithium sublattice. At 600~K, Li-ions remain ordered and essentially immobile, consistent with the tetragonal symmetry. Between 800 and 1000~K, however, the mean-squared displacement of Li ions begins to increase sharply, signaling the onset of Li delocalization and dynamic disorder, while the host La--Zr--O framework remains intact. This mechanism is in line with experimental observations: Geiger \textit{et al.}~\cite{Geiger2011} reported that aliovalent doping or Li deficiency stabilizes the cubic phase of LLZO down to 100--150~$^\circ$C, where partially occupied Li sites lead to enhanced ionic conductivity. In contrast, Matsui \textit{et al.}~\cite{Matsui2014} found that stoichiometric, Al-free LLZO retains the ordered tetragonal structure up to much higher temperatures ($\sim$650~$^\circ$C), with the high-temperature cubic phase appearing only when additional Li-vacancies are introduced or CO$_2$ uptake promotes Li-site disorder. These results together indicate that the t--c transition temperature depends sensitively on Li-vacancy concentration and sample chemistry, reconciling the higher intrinsic transition temperature of pure LLZO with the lower values observed in doped or vacancy-rich compositions.

The coupling between Li sublattice disorder and structural symmetry is also strongly supported by computational studies \cite{Hong2024LLZO_PhaseDiagram}. Bernstein \textit{et al.}~\cite{Bernstein2012} used DFT and AIMD to show that the t--c transition is an order--disorder process of the Li sublattice, with vacancy-induced entropy gain stabilizing the cubic form. Meier \textit{et al.}~\cite{Meier2014} further demonstrated that Li diffusion mechanisms change fundamentally between the two phases: collective, synchronous Li motion dominates in tetragonal LLZO, whereas single-ion asynchronous hopping is enabled in cubic LLZO, lowering activation barriers. Neutron scattering and reverse Monte Carlo studies confirm this picture, showing that only a fraction of Li ions are mobile at a given time in the cubic phase, but over long times all participate in diffusion, consistent with a partially melted Li sublattice~\cite{Klenk2015,Tian2023}. Our present simulations, which capture both the lattice constant convergence ($a=b=c$) and the rapid rise in Li diffusivity near 1000~K, provide direct evidence that the t--c transition is a thermally activated superionic transformation of the lithium sublattice, fully consistent with experimental and theoretical benchmarks.

\begin{figure}[t]
  \centering
  \includegraphics[width=0.16\textwidth]{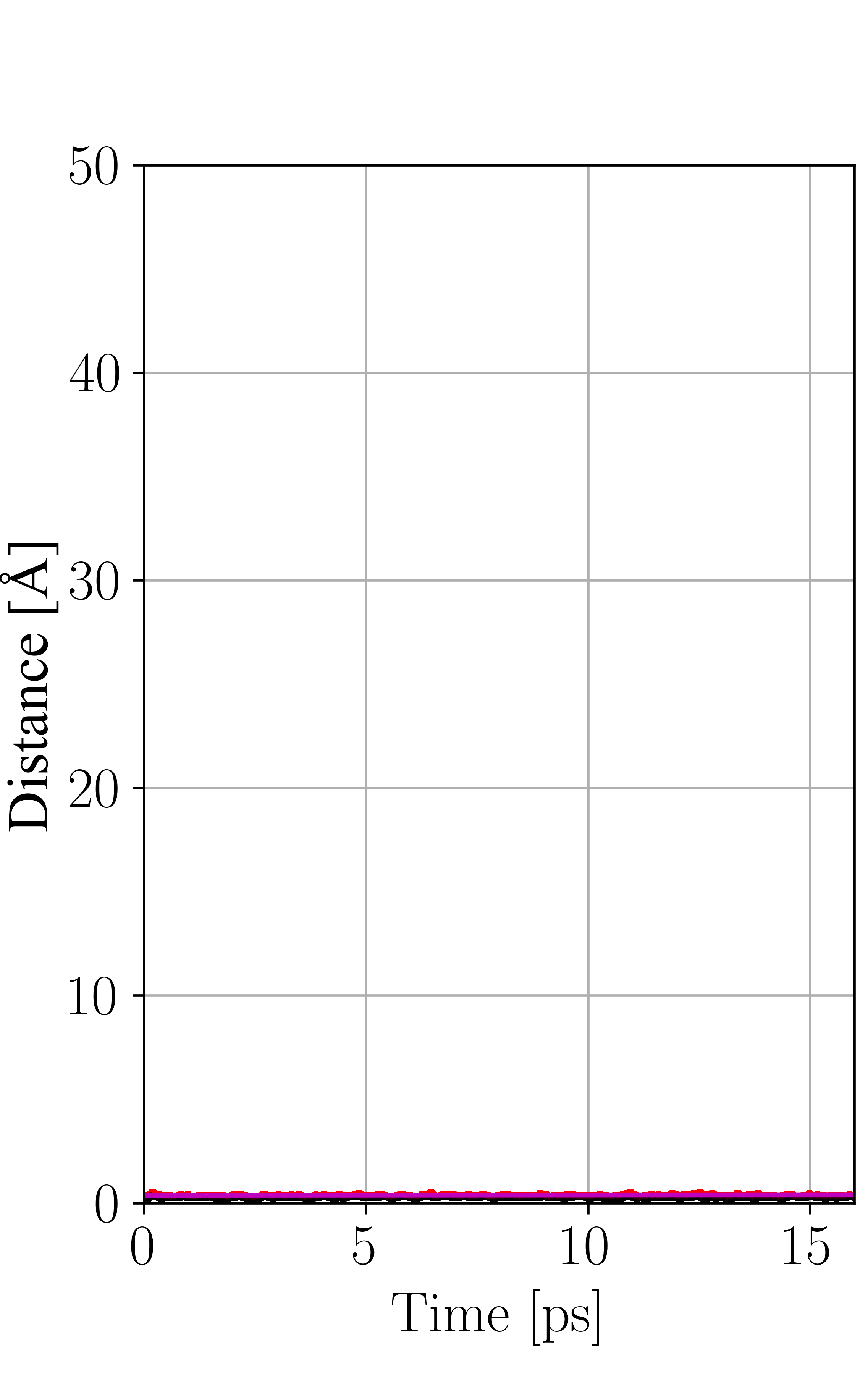}\hfill
  \includegraphics[width=0.16\textwidth]{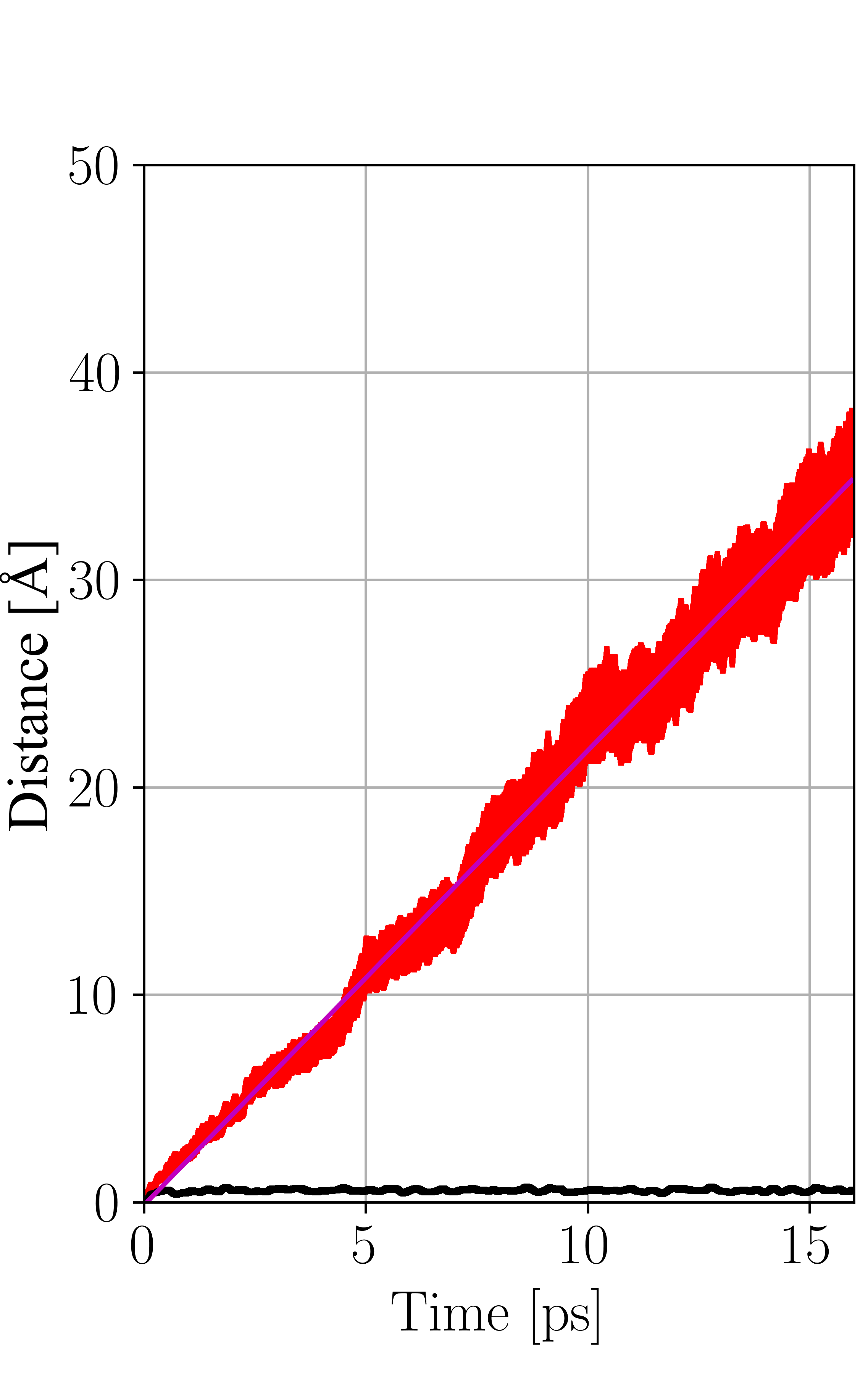}\hfill
  \includegraphics[width=0.16\textwidth]{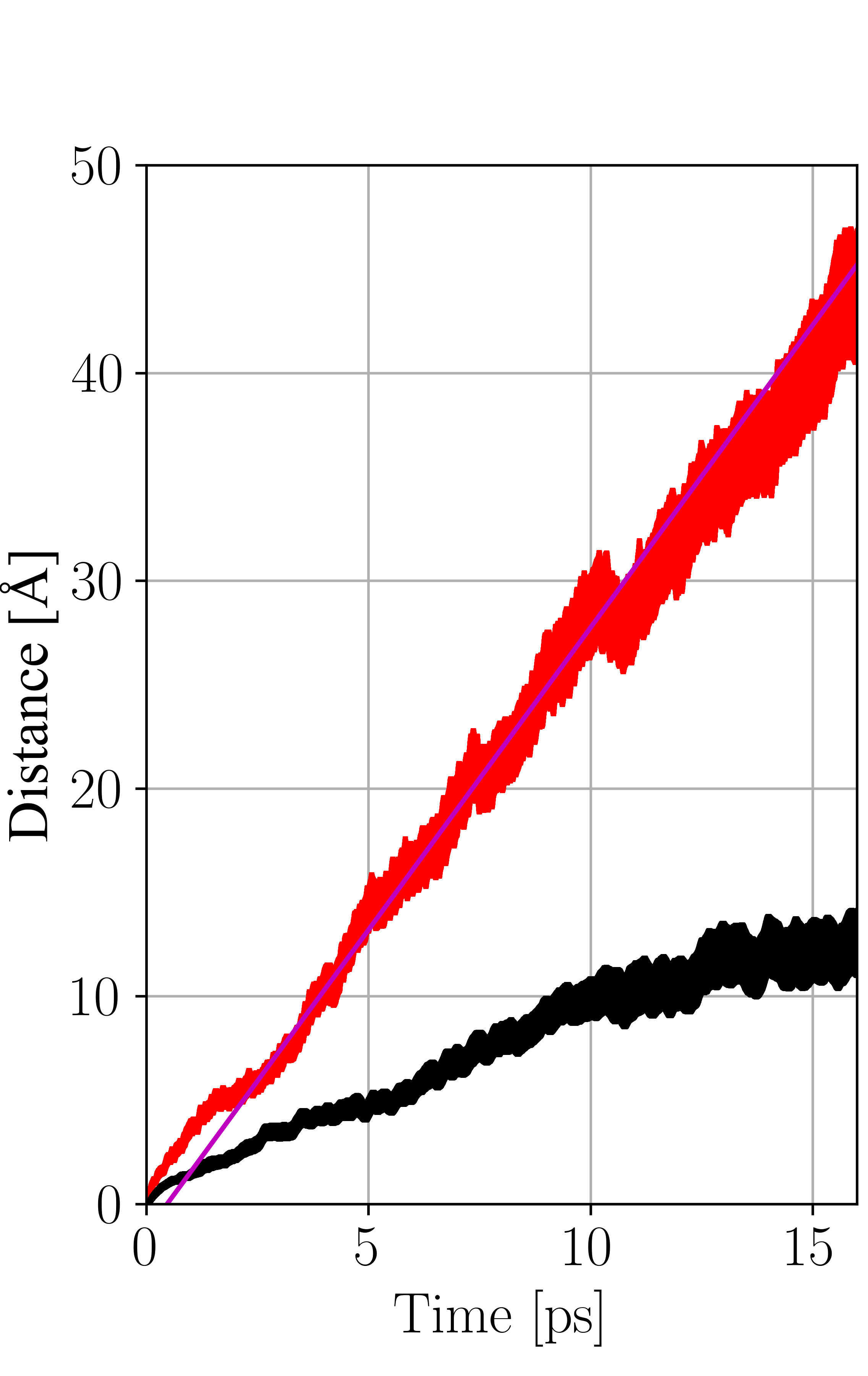}
  \caption{Oxygen diffusion across Er$_2$O$_3$ phases at 2700~K. 
  Left: {C phase} (solid): O MSD remains flat; $D_\mathrm{O}\!\approx\!0$.
  Middle: {H phase}: linear O MSD with finite slope while cations remain nearly immobile, evidencing oxygen sublattice melting (superionic-like behavior).
  Right: {L phase} (liquid): large linear MSD and high diffusivity for all species. Linear fits (purple trace) indicate the diffusive windows used for Einstein-slope evaluation.}
  \label{fig:er2o3_diffusion_panels}
\end{figure}

\subsubsection{Oxygen Diffusion and Sublattice Melting in Er$_2$O$_3$}

AIMD reveals three distinct transport regimes for oxygen in Er$_2$O$_3$ that correlate with the crystalline/liquid state of the host lattice. In the {C phase} (fully solid), the MSD of O atoms remains nearly flat over the production window and the linear MSD fit yields a diffusivity indistinguishable from zero within uncertainty, consistent with arrested oxygen dynamics in a rigid framework. In sharp contrast, the {L phase} (fully molten) exhibits a large, strictly linear MSD with time, producing diffusion coefficients characteristic of high-temperature oxide melts; oxygen, erbium, and the network oxygens move collectively, and the slope is stable across fitting windows, indicating well-developed Brownian diffusion. 

The {H phase} sits between these limits and shows the hallmark of {oxygen sublattice melting}, consistent with our previous first-principles discovery of the same phenomenon in the H phase of Y$_2$O$_3$ \cite{Wang2024Acta,Ushakov2024Er2O3_Tm2O3}, cubic ZrO$_2$, and hafnium carbonitride \cite{Hong2015HfC_TaC_Melting,HongLiu2025Entropy}. While the cation framework retains substantial crystalline order, the O-only MSD becomes linear with a pronounced slope, yielding a finite self-diffusion coefficient comparable to the liquid. This decoupling—mobile oxygen on a comparatively rigid cation scaffold—indicates partial disordering (``melting'') of the oxygen sublattice without wholesale loss of long-range order in the cation lattice. Practically, the H phase displays (i) a clear separation between $D_\mathrm{O}$ and the near-zero cation diffusivities, (ii) an MSD linear regime that emerges after a short transient, and (iii) robustness of the Einstein-slope estimate across reasonable fitting windows, all consistent with a superionic-like state localized to the oxygen network.

\subsection*{Oxygen diffusion in bcc and fcc Fe (solids) and in FeO liquids and the impact of SiO$_2$ and Al$_2$O$_3$}

\begin{figure}[t]
  \centering
  \includegraphics[width=0.49\textwidth]{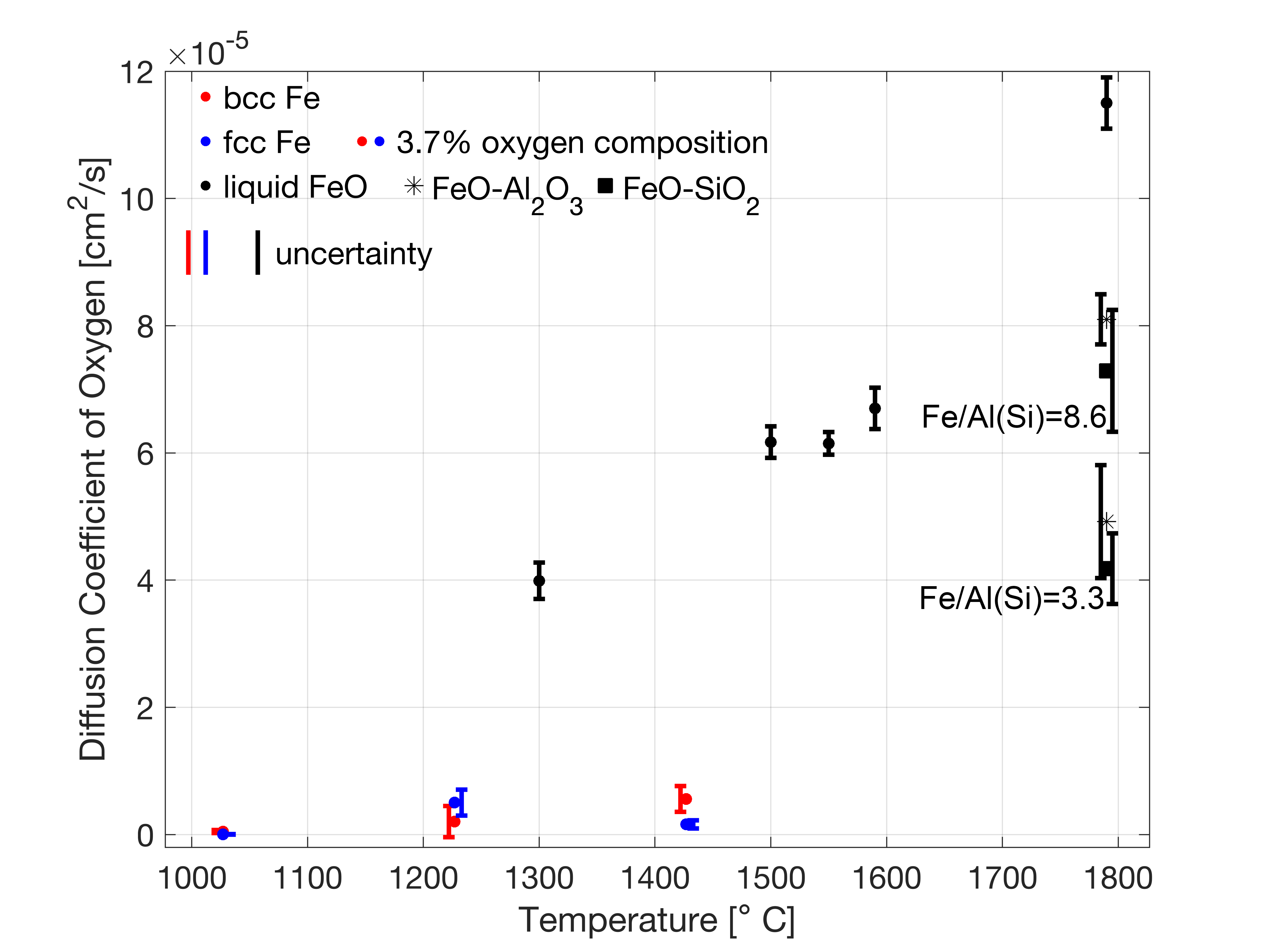} 
  \caption{Calculated diffusion coefficient of oxygen in bcc and fcc iron at 1027, 1227, and 1427 °C (1300, 1500, and 1700 K) for oxygen concentration of 3.7 atomic \%, and computed values of diffusion coefficient of O in liquid FeO at different temperatures. Uncertainties are determined based on four segments of molecular dynamic simulations for each temperature, phase, and composition.}
  \label{fig:O_in_FeO}
\end{figure}

Oxygen mobility in iron strongly depends on phase state. In the solid, interstitial oxygen diffuses more readily in $\alpha$-Fe (bcc) than in $\gamma$-Fe (fcc). Internal oxidation experiments have shown that the activation energy for O diffusion is ~96~kJ/mol in bcc iron, compared to 166~kJ/mol in fcc iron~\cite{Takada1986-Alpha,Takada1986-Gamma}. As a result, AIMD simulations confirm nearly an order-of-magnitude difference: at 1700~K, $D_{\mathrm{O}} \approx 5.6\times10^{-6}$~cm$^2$/s in bcc Fe, but only $1.6\times10^{-6}$~cm$^2$/s in fcc Fe. The sluggish oxygen transport in $\gamma$-Fe effectively ``locks'' oxygen in the lattice and, together with its inherently low permeability, is a key factor behind the slowed reduction kinetics observed in the austenitic regime \cite{Mohanta2025HydrogenReduction}.

Once the FeO oxide melts, oxygen diffusivity increases dramatically. In molten wüstite (FeO), tracer experiments report $D_{\mathrm{O}}\sim10^{-3}$--$10^{-4}$~cm$^2$/s near 1900~K~\cite{Dunn1983}. Our AIMD calculations similarly show a steady increase in oxygen diffusivity between 1300 and 1800~$^\circ$C \cite{Mohanta2025MeltThenReduce}. The addition of aluminum and silicon further modifies transport: in FeO–Al$_2$O$_3$ and FeO–SiO$_2$ liquids, O diffusivity decreases with increasing Si and Al fractions due to network polymerization, as confirmed by geochemical diffusion experiments in silicate melts~\cite{Dunn1986-ChemGeol}. Nonetheless, simulations at Earth-core–like conditions show that oxygen remains the fastest diffusing species in Fe--Si--O liquids, typically 2--3$\times$ faster than Fe or Si self-diffusion~\cite{Pozzo2013}. 

Together, these results establish a coherent picture: oxygen diffuses quickly in bcc Fe, sluggishly in fcc Fe, and extremely rapidly in FeO liquids, with Si and Al concentration serving as a key control knob for oxygen mobility in molten alloys. This behavior arises because Si and Al in the melt do not exist as isolated cations but form extended polymeric complexes (e.g., SiO$_4$ and AlO$_4$ tetrahedra) that are linked through shared oxygen atoms. These network-forming units significantly increase the degree of melt polymerization, thereby reducing oxygen mobility and enhancing viscous resistance.

It should be clarified that the simulations represent an oxide melt—chemically corresponding to an FeO–Al$_2$O$_3$–SiO$_2$ system—rather than a metallic melt with dissolved oxygen and silicon. In such ionic liquids, Fe primarily exists as Fe$^{2+}$ cations and O$^{2-}$ as anions, while Si and Al form polymerized network units (SiO$_4$ and AlO$_4$) whose connectivity depends on the melt basicity (i.e., the FeO/(SiO$_2$+Al$_2$O$_3$) ratio). Because of this polymeric structure, transport cannot be fully described by the simple Stokes–Einstein relation, and deviations between diffusion and viscosity are expected.

\section{Conclusions}

We have extended \texttt{SLUSCHI} from its original role in first-principles melting calculations to a fully automated pipeline for computing diffusion in liquids and in crystalline solids exhibiting sublattice melting. The workflow isolates the thermal-expansion/volume-search stage to equilibrate the supercell at the target $(T,P)$, generates production AIMD trajectories, and post-processes \texttt{VASP} outputs to extract species-resolved mean-squared displacements and self-diffusion coefficients via the Einstein relation, with block-averaged uncertainty and diagnostic plots (MSD, running slopes, VACF) for quality control. Validation across representative cases—Al–Cu liquids (linking $D$ and $\eta$ through Stokes–Einstein; see Table~II), lithium sublattice melting in LLZO (Fig.~2), oxygen sublattice melting in $\mathrm{Er_2O_3}$ (Fig.~6), and oxygen mobility in bcc/fcc Fe and FeO–Al$_2$O$_3$–SiO$_2$ liquids (Fig.~7)—demonstrates that the approach captures high-mobility regimes with quantitative fidelity while retaining the robustness and ease-of-use that characterize \texttt{SLUSCHI}. 

Because it reuses \texttt{SLUSCHI}'s automation (input generation, job orchestration, and structured outputs) but terminates before the melt/coexistence steps, the diffusion module is both portable and high-throughput: only the opt/volume-search stages are required, and practical accelerators (e.g., the special $(\tfrac14,\tfrac14,\tfrac14)$ $k$-point option) can lower cost without sacrificing accuracy for targeted properties. Looking ahead, ensemble averaging over multiple seeds, modest finite-size scaling via the \texttt{radius} tag, optional Green–Kubo viscosity alongside Einstein diffusivities, and tighter hooks to CALPHAD assessments will further strengthen predictive transport databases; for light ions or quantum liquids, path-integral MD is a natural extension, and active-learning interatomic potentials can extend length/time scales while preserving DFT reference quality. Taken together, these elements position “\texttt{SLUSCHI}–Diffusion” as a practical, reproducible route to first-principles transport data across metallic and oxide systems relevant to high-temperature processing and materials design. 

\section*{Acknowledgments}
This research was supported by U.S. Department of Energy (DOE), Office of Science, Office of Basic Energy Sciences (BES), Materials Sciences and Engineering Division under Award DE-SC0024724 “\textit{Fundamental Studies of Hydrogen Arc Plasmas for High-efficiency and Carbon-free Steelmaking}”, and by Arizona State University through the use of the facilities at its Research Computing. This work was authored in part by National Renewable Energy Laboratory (NREL), operated under Contract No. DE-AC36-08GO28308. The views expressed in the article do not necessarily represent the views of the DOE or the U.S. Government. The U.S. Government retains and the publisher, by accepting the article for publication, acknowledges that the U.S. Government retains a nonexclusive, paid-up, irrevocable, worldwide license to publish or reproduce the published form of this work, or allow others to do so, for U.S. Government purposes.

\section*{Data Availability}
The data that support the findings of this study are available from the corresponding author upon reasonable request. Input and output files for representative SLUSCHI diffusion and viscosity calculations are also provided as example datasets in the SLUSCHI GitHub repository (https://github.com/qjhong/sluschi).

\bibliography{references} 

\begin{thebibliography}{43}%
\makeatletter
\providecommand \@ifxundefined [1]{%
 \@ifx{#1\undefined}
}%
\providecommand \@ifnum [1]{%
 \ifnum #1\expandafter \@firstoftwo
 \else \expandafter \@secondoftwo
 \fi
}%
\providecommand \@ifx [1]{%
 \ifx #1\expandafter \@firstoftwo
 \else \expandafter \@secondoftwo
 \fi
}%
\providecommand \natexlab [1]{#1}%
\providecommand \enquote  [1]{``#1''}%
\providecommand \bibnamefont  [1]{#1}%
\providecommand \bibfnamefont [1]{#1}%
\providecommand \citenamefont [1]{#1}%
\providecommand \href@noop [0]{\@secondoftwo}%
\providecommand \href [0]{\begingroup \@sanitize@url \@href}%
\providecommand \@href[1]{\@@startlink{#1}\@@href}%
\providecommand \@@href[1]{\endgroup#1\@@endlink}%
\providecommand \@sanitize@url [0]{\catcode `\\12\catcode `\$12\catcode
  `\&12\catcode `\#12\catcode `\^12\catcode `\_12\catcode `\%12\relax}%
\providecommand \@@startlink[1]{}%
\providecommand \@@endlink[0]{}%
\providecommand \url  [0]{\begingroup\@sanitize@url \@url }%
\providecommand \@url [1]{\endgroup\@href {#1}{\urlprefix }}%
\providecommand \urlprefix  [0]{URL }%
\providecommand \Eprint [0]{\href }%
\providecommand \doibase [0]{https://doi.org/}%
\providecommand \selectlanguage [0]{\@gobble}%
\providecommand \bibinfo  [0]{\@secondoftwo}%
\providecommand \bibfield  [0]{\@secondoftwo}%
\providecommand \translation [1]{[#1]}%
\providecommand \BibitemOpen [0]{}%
\providecommand \bibitemStop [0]{}%
\providecommand \bibitemNoStop [0]{.\EOS\space}%
\providecommand \EOS [0]{\spacefactor3000\relax}%
\providecommand \BibitemShut  [1]{\csname bibitem#1\endcsname}%
\let\auto@bib@innerbib\@empty
\bibitem [{\citenamefont {Sandberg}\ \emph {et~al.}(2002)\citenamefont
  {Sandberg}, \citenamefont {Magyari-K{\o}pe},\ and\ \citenamefont
  {Mattsson}}]{Sandberg2002PRL}%
  \BibitemOpen
  \bibfield  {author} {\bibinfo {author} {\bibfnamefont {N.}~\bibnamefont
  {Sandberg}}, \bibinfo {author} {\bibfnamefont {B.}~\bibnamefont
  {Magyari-K{\o}pe}},\ and\ \bibinfo {author} {\bibfnamefont {T.~R.}\
  \bibnamefont {Mattsson}},\ }\bibfield  {title} {\bibinfo {title}
  {Self-diffusion rates in al from combined first-principles and
  model-potential calculations},\ }\href
  {https://doi.org/10.1103/PhysRevLett.89.065901} {\bibfield  {journal}
  {\bibinfo  {journal} {Physical Review Letters}\ }\textbf {\bibinfo {volume}
  {89}},\ \bibinfo {pages} {065901} (\bibinfo {year} {2002})}\BibitemShut
  {NoStop}%
\bibitem [{\citenamefont {Mantina}\ \emph {et~al.}(2008)\citenamefont
  {Mantina}, \citenamefont {Wang}, \citenamefont {Arroyave}, \citenamefont
  {Chen}, \citenamefont {Liu},\ and\ \citenamefont
  {Wolverton}}]{Mantina2008PRL}%
  \BibitemOpen
  \bibfield  {author} {\bibinfo {author} {\bibfnamefont {M.}~\bibnamefont
  {Mantina}}, \bibinfo {author} {\bibfnamefont {Y.}~\bibnamefont {Wang}},
  \bibinfo {author} {\bibfnamefont {R.}~\bibnamefont {Arroyave}}, \bibinfo
  {author} {\bibfnamefont {L.-Q.}\ \bibnamefont {Chen}}, \bibinfo {author}
  {\bibfnamefont {Z.-K.}\ \bibnamefont {Liu}},\ and\ \bibinfo {author}
  {\bibfnamefont {C.}~\bibnamefont {Wolverton}},\ }\bibfield  {title} {\bibinfo
  {title} {First-principles calculation of self-diffusion coefficients},\
  }\href {https://doi.org/10.1103/PhysRevLett.100.215901} {\bibfield  {journal}
  {\bibinfo  {journal} {Physical Review Letters}\ }\textbf {\bibinfo {volume}
  {100}},\ \bibinfo {pages} {215901} (\bibinfo {year} {2008})}\BibitemShut
  {NoStop}%
\bibitem [{\citenamefont {Mantina}\ \emph {et~al.}(2009)\citenamefont
  {Mantina}, \citenamefont {Wang}, \citenamefont {Chen}, \citenamefont {Liu},\
  and\ \citenamefont {Wolverton}}]{Mantina2009Acta}%
  \BibitemOpen
  \bibfield  {author} {\bibinfo {author} {\bibfnamefont {M.}~\bibnamefont
  {Mantina}}, \bibinfo {author} {\bibfnamefont {Y.}~\bibnamefont {Wang}},
  \bibinfo {author} {\bibfnamefont {L.-Q.}\ \bibnamefont {Chen}}, \bibinfo
  {author} {\bibfnamefont {Z.-K.}\ \bibnamefont {Liu}},\ and\ \bibinfo {author}
  {\bibfnamefont {C.}~\bibnamefont {Wolverton}},\ }\bibfield  {title} {\bibinfo
  {title} {First-principles impurity diffusion coefficients},\ }\href
  {https://doi.org/10.1016/j.actamat.2009.05.006} {\bibfield  {journal}
  {\bibinfo  {journal} {Acta Materialia}\ }\textbf {\bibinfo {volume} {57}},\
  \bibinfo {pages} {4102} (\bibinfo {year} {2009})}\BibitemShut {NoStop}%
\bibitem [{\citenamefont {Bl\"ochl}\ \emph {et~al.}(1993)\citenamefont
  {Bl\"ochl}, \citenamefont {Smargiassi}, \citenamefont {Car}, \citenamefont
  {Laks}, \citenamefont {Andreoni},\ and\ \citenamefont
  {Pantelides}}]{Blochl1993PRL}%
  \BibitemOpen
  \bibfield  {author} {\bibinfo {author} {\bibfnamefont {P.~E.}\ \bibnamefont
  {Bl\"ochl}}, \bibinfo {author} {\bibfnamefont {E.}~\bibnamefont
  {Smargiassi}}, \bibinfo {author} {\bibfnamefont {R.}~\bibnamefont {Car}},
  \bibinfo {author} {\bibfnamefont {D.~B.}\ \bibnamefont {Laks}}, \bibinfo
  {author} {\bibfnamefont {W.}~\bibnamefont {Andreoni}},\ and\ \bibinfo
  {author} {\bibfnamefont {S.~T.}\ \bibnamefont {Pantelides}},\ }\bibfield
  {title} {\bibinfo {title} {First-principles calculations of self-diffusion
  constants in silicon},\ }\href {https://doi.org/10.1103/PhysRevLett.70.2435}
  {\bibfield  {journal} {\bibinfo  {journal} {Physical Review Letters}\
  }\textbf {\bibinfo {volume} {70}},\ \bibinfo {pages} {2435} (\bibinfo {year}
  {1993})}\BibitemShut {NoStop}%
\bibitem [{\citenamefont {Milman}\ \emph {et~al.}(1993)\citenamefont {Milman},
  \citenamefont {Payne}, \citenamefont {Heine}, \citenamefont {Needs},
  \citenamefont {Lin},\ and\ \citenamefont {Lee}}]{Milman1993PRL}%
  \BibitemOpen
  \bibfield  {author} {\bibinfo {author} {\bibfnamefont {V.}~\bibnamefont
  {Milman}}, \bibinfo {author} {\bibfnamefont {M.~C.}\ \bibnamefont {Payne}},
  \bibinfo {author} {\bibfnamefont {V.}~\bibnamefont {Heine}}, \bibinfo
  {author} {\bibfnamefont {R.~J.}\ \bibnamefont {Needs}}, \bibinfo {author}
  {\bibfnamefont {J.~S.}\ \bibnamefont {Lin}},\ and\ \bibinfo {author}
  {\bibfnamefont {M.~H.}\ \bibnamefont {Lee}},\ }\bibfield  {title} {\bibinfo
  {title} {Free energy and entropy of diffusion by ab initio molecular
  dynamics: Alkali ions in silicon},\ }\href
  {https://doi.org/10.1103/PhysRevLett.70.2928} {\bibfield  {journal} {\bibinfo
   {journal} {Physical Review Letters}\ }\textbf {\bibinfo {volume} {70}},\
  \bibinfo {pages} {2928} (\bibinfo {year} {1993})}\BibitemShut {NoStop}%
\bibitem [{\citenamefont {Henkelman}\ \emph {et~al.}(2000)\citenamefont
  {Henkelman}, \citenamefont {Uberuaga},\ and\ \citenamefont
  {J{\'o}nsson}}]{Henkelman2000CI}%
  \BibitemOpen
  \bibfield  {author} {\bibinfo {author} {\bibfnamefont {G.}~\bibnamefont
  {Henkelman}}, \bibinfo {author} {\bibfnamefont {B.~P.}\ \bibnamefont
  {Uberuaga}},\ and\ \bibinfo {author} {\bibfnamefont {H.}~\bibnamefont
  {J{\'o}nsson}},\ }\bibfield  {title} {\bibinfo {title} {A climbing image
  nudged elastic band method for finding saddle points and minimum energy
  paths},\ }\href {https://doi.org/10.1063/1.1329672} {\bibfield  {journal}
  {\bibinfo  {journal} {The Journal of Chemical Physics}\ }\textbf {\bibinfo
  {volume} {113}},\ \bibinfo {pages} {9901} (\bibinfo {year}
  {2000})}\BibitemShut {NoStop}%
\bibitem [{\citenamefont {Henkelman}\ and\ \citenamefont
  {J{\'o}nsson}(2000)}]{Henkelman2000NEB}%
  \BibitemOpen
  \bibfield  {author} {\bibinfo {author} {\bibfnamefont {G.}~\bibnamefont
  {Henkelman}}\ and\ \bibinfo {author} {\bibfnamefont {H.}~\bibnamefont
  {J{\'o}nsson}},\ }\bibfield  {title} {\bibinfo {title} {Improved tangent
  estimate in the nudged elastic band method for finding minimum energy paths
  and saddle points},\ }\href {https://doi.org/10.1063/1.1323224} {\bibfield
  {journal} {\bibinfo  {journal} {The Journal of Chemical Physics}\ }\textbf
  {\bibinfo {volume} {113}},\ \bibinfo {pages} {9978} (\bibinfo {year}
  {2000})}\BibitemShut {NoStop}%
\bibitem [{\citenamefont {Henkelman}\ and\ \citenamefont
  {J{\'o}nsson}(1999)}]{Henkelman1999Dimer}%
  \BibitemOpen
  \bibfield  {author} {\bibinfo {author} {\bibfnamefont {G.}~\bibnamefont
  {Henkelman}}\ and\ \bibinfo {author} {\bibfnamefont {H.}~\bibnamefont
  {J{\'o}nsson}},\ }\bibfield  {title} {\bibinfo {title} {A dimer method for
  finding saddle points on high dimensional potential surfaces using only first
  derivatives},\ }\href {https://doi.org/10.1063/1.480097} {\bibfield
  {journal} {\bibinfo  {journal} {The Journal of Chemical Physics}\ }\textbf
  {\bibinfo {volume} {111}},\ \bibinfo {pages} {7010} (\bibinfo {year}
  {1999})}\BibitemShut {NoStop}%
\bibitem [{\citenamefont {Mayeshiba}\ \emph {et~al.}(2017)\citenamefont
  {Mayeshiba}, \citenamefont {Wu}, \citenamefont {Angsten}, \citenamefont
  {Kaczmarowski}, \citenamefont {Song}, \citenamefont {Jenness}, \citenamefont
  {Xie},\ and\ \citenamefont {Morgan}}]{Mayeshiba2017MAST}%
  \BibitemOpen
  \bibfield  {author} {\bibinfo {author} {\bibfnamefont {T.}~\bibnamefont
  {Mayeshiba}}, \bibinfo {author} {\bibfnamefont {H.}~\bibnamefont {Wu}},
  \bibinfo {author} {\bibfnamefont {T.}~\bibnamefont {Angsten}}, \bibinfo
  {author} {\bibfnamefont {A.}~\bibnamefont {Kaczmarowski}}, \bibinfo {author}
  {\bibfnamefont {Z.}~\bibnamefont {Song}}, \bibinfo {author} {\bibfnamefont
  {G.}~\bibnamefont {Jenness}}, \bibinfo {author} {\bibfnamefont
  {W.}~\bibnamefont {Xie}},\ and\ \bibinfo {author} {\bibfnamefont
  {D.}~\bibnamefont {Morgan}},\ }\bibfield  {title} {\bibinfo {title} {The
  materials simulation toolkit (mast) for atomistic modeling of defects and
  diffusion},\ }\href@noop {} {\bibfield  {journal} {\bibinfo  {journal}
  {Computational Materials Science}\ }\textbf {\bibinfo {volume} {126}},\
  \bibinfo {pages} {90} (\bibinfo {year} {2017})}\BibitemShut {NoStop}%
\bibitem [{\citenamefont {Mathew}\ \emph {et~al.}(2017)\citenamefont {Mathew},
  \citenamefont {Montoya}, \citenamefont {Faghaninia}, \citenamefont
  {Dwarakanath}, \citenamefont {Aykol}, \citenamefont {Tang}, \citenamefont
  {Chu}, \citenamefont {Smidt}, \citenamefont {Bocklund}, \citenamefont
  {Horton}, \citenamefont {Dagdelen}, \citenamefont {Wood}, \citenamefont
  {Liu}, \citenamefont {Neaton}, \citenamefont {Ong}, \citenamefont {Persson},\
  and\ \citenamefont {Jain}}]{Mathew2017Atomate}%
  \BibitemOpen
  \bibfield  {author} {\bibinfo {author} {\bibfnamefont {K.}~\bibnamefont
  {Mathew}}, \bibinfo {author} {\bibfnamefont {J.~H.}\ \bibnamefont {Montoya}},
  \bibinfo {author} {\bibfnamefont {A.}~\bibnamefont {Faghaninia}}, \bibinfo
  {author} {\bibfnamefont {S.}~\bibnamefont {Dwarakanath}}, \bibinfo {author}
  {\bibfnamefont {M.}~\bibnamefont {Aykol}}, \bibinfo {author} {\bibfnamefont
  {H.}~\bibnamefont {Tang}}, \bibinfo {author} {\bibfnamefont {I.-H.}\
  \bibnamefont {Chu}}, \bibinfo {author} {\bibfnamefont {T.}~\bibnamefont
  {Smidt}}, \bibinfo {author} {\bibfnamefont {B.}~\bibnamefont {Bocklund}},
  \bibinfo {author} {\bibfnamefont {M.}~\bibnamefont {Horton}}, \bibinfo
  {author} {\bibfnamefont {J.}~\bibnamefont {Dagdelen}}, \bibinfo {author}
  {\bibfnamefont {B.}~\bibnamefont {Wood}}, \bibinfo {author} {\bibfnamefont
  {Z.-K.}\ \bibnamefont {Liu}}, \bibinfo {author} {\bibfnamefont
  {J.}~\bibnamefont {Neaton}}, \bibinfo {author} {\bibfnamefont {S.~P.}\
  \bibnamefont {Ong}}, \bibinfo {author} {\bibfnamefont {K.~A.}\ \bibnamefont
  {Persson}},\ and\ \bibinfo {author} {\bibfnamefont {A.}~\bibnamefont
  {Jain}},\ }\bibfield  {title} {\bibinfo {title} {Atomate: A high-level
  interface to generate, execute, and analyze computational materials science
  workflows},\ }\href {https://doi.org/10.1016/j.commatsci.2017.07.030}
  {\bibfield  {journal} {\bibinfo  {journal} {Computational Materials Science}\
  }\textbf {\bibinfo {volume} {139}},\ \bibinfo {pages} {140} (\bibinfo {year}
  {2017})}\BibitemShut {NoStop}%
\bibitem [{\citenamefont {contributors}(2023)}]{MPmorph2023}%
  \BibitemOpen
  \bibfield  {author} {\bibinfo {author} {\bibfnamefont {M.}~\bibnamefont
  {contributors}},\ }\href@noop {} {\bibinfo {title} {Mpmorph: Aimd workflows
  and analysis tools (github)}},\ \bibinfo {howpublished}
  {\url{https://github.com/materialsproject/mpmorph}} (\bibinfo {year}
  {2023})\BibitemShut {NoStop}%
\bibitem [{\citenamefont {Larsen}\ \emph {et~al.}(2017)\citenamefont {Larsen},
  \citenamefont {Mortensen}, \citenamefont {Blomqvist}, \citenamefont
  {Castelli}, \citenamefont {Christensen}, \citenamefont {Du{\l}ak},
  \citenamefont {Friis}, \citenamefont {Groves}, \citenamefont {Hammer},
  \citenamefont {Hargus}, \citenamefont {Hermes}, \citenamefont {Jennings},
  \citenamefont {Jensen}, \citenamefont {Kermode}, \citenamefont {Kitchin},
  \citenamefont {Kolsbjerg}, \citenamefont {Kubal}, \citenamefont {Kaasbjerg},
  \citenamefont {Lysgaard}, \citenamefont {Maronsson}, \citenamefont {Maxson},
  \citenamefont {Olsen}, \citenamefont {Pastewka}, \citenamefont {Peterson},
  \citenamefont {Rostgaard}, \citenamefont {Sch{\'a}fer}, \citenamefont
  {Thygesen}, \citenamefont {Tsukerman}, \citenamefont {Vegge}, \citenamefont
  {Vilhelmsen}, \citenamefont {Walter}, \citenamefont {Zeng},\ and\
  \citenamefont {Jacobsen}}]{Larsen2017ASE}%
  \BibitemOpen
  \bibfield  {author} {\bibinfo {author} {\bibfnamefont {A.~H.}\ \bibnamefont
  {Larsen}}, \bibinfo {author} {\bibfnamefont {J.~J.}\ \bibnamefont
  {Mortensen}}, \bibinfo {author} {\bibfnamefont {J.}~\bibnamefont
  {Blomqvist}}, \bibinfo {author} {\bibfnamefont {I.~E.}\ \bibnamefont
  {Castelli}}, \bibinfo {author} {\bibfnamefont {R.}~\bibnamefont
  {Christensen}}, \bibinfo {author} {\bibfnamefont {M.}~\bibnamefont
  {Du{\l}ak}}, \bibinfo {author} {\bibfnamefont {J.}~\bibnamefont {Friis}},
  \bibinfo {author} {\bibfnamefont {M.~N.}\ \bibnamefont {Groves}}, \bibinfo
  {author} {\bibfnamefont {B.}~\bibnamefont {Hammer}}, \bibinfo {author}
  {\bibfnamefont {C.}~\bibnamefont {Hargus}}, \bibinfo {author} {\bibfnamefont
  {E.~D.}\ \bibnamefont {Hermes}}, \bibinfo {author} {\bibfnamefont {P.~C.}\
  \bibnamefont {Jennings}}, \bibinfo {author} {\bibfnamefont {P.~B.}\
  \bibnamefont {Jensen}}, \bibinfo {author} {\bibfnamefont {J.}~\bibnamefont
  {Kermode}}, \bibinfo {author} {\bibfnamefont {J.~R.}\ \bibnamefont
  {Kitchin}}, \bibinfo {author} {\bibfnamefont {E.~L.}\ \bibnamefont
  {Kolsbjerg}}, \bibinfo {author} {\bibfnamefont {J.}~\bibnamefont {Kubal}},
  \bibinfo {author} {\bibfnamefont {K.}~\bibnamefont {Kaasbjerg}}, \bibinfo
  {author} {\bibfnamefont {S.}~\bibnamefont {Lysgaard}}, \bibinfo {author}
  {\bibfnamefont {J.~B.}\ \bibnamefont {Maronsson}}, \bibinfo {author}
  {\bibfnamefont {T.}~\bibnamefont {Maxson}}, \bibinfo {author} {\bibfnamefont
  {T.}~\bibnamefont {Olsen}}, \bibinfo {author} {\bibfnamefont
  {L.}~\bibnamefont {Pastewka}}, \bibinfo {author} {\bibfnamefont
  {A.}~\bibnamefont {Peterson}}, \bibinfo {author} {\bibfnamefont
  {C.}~\bibnamefont {Rostgaard}}, \bibinfo {author} {\bibfnamefont
  {J.}~\bibnamefont {Sch{\'a}fer}}, \bibinfo {author} {\bibfnamefont {K.~S.}\
  \bibnamefont {Thygesen}}, \bibinfo {author} {\bibfnamefont {I.~P.}\
  \bibnamefont {Tsukerman}}, \bibinfo {author} {\bibfnamefont {M.}~\bibnamefont
  {Vegge}}, \bibinfo {author} {\bibfnamefont {F.}~\bibnamefont {Vilhelmsen}},
  \bibinfo {author} {\bibfnamefont {M.}~\bibnamefont {Walter}}, \bibinfo
  {author} {\bibfnamefont {Z.}~\bibnamefont {Zeng}},\ and\ \bibinfo {author}
  {\bibfnamefont {K.~W.}\ \bibnamefont {Jacobsen}},\ }\bibfield  {title}
  {\bibinfo {title} {The atomic simulation environment — a python library for
  working with atoms},\ }\href {https://doi.org/10.1088/1361-648X/aa680e}
  {\bibfield  {journal} {\bibinfo  {journal} {Journal of Physics: Condensed
  Matter}\ }\textbf {\bibinfo {volume} {29}},\ \bibinfo {pages} {273002}
  (\bibinfo {year} {2017})}\BibitemShut {NoStop}%
\bibitem [{\citenamefont {Smidstrup}\ \emph {et~al.}(2020)\citenamefont
  {Smidstrup}, \citenamefont {Markussen}, \citenamefont {Stokbro},\ and\
  \citenamefont {et~al.}}]{Smidstrup2020QuantumATK}%
  \BibitemOpen
  \bibfield  {author} {\bibinfo {author} {\bibfnamefont {S.}~\bibnamefont
  {Smidstrup}}, \bibinfo {author} {\bibfnamefont {A.}~\bibnamefont
  {Markussen}}, \bibinfo {author} {\bibfnamefont {K.~A.}\ \bibnamefont
  {Stokbro}},\ and\ \bibinfo {author} {\bibnamefont {et~al.}},\ }\bibfield
  {title} {\bibinfo {title} {Quantumatk: An integrated platform of electronic
  and atomic-scale modelling tools},\ }\href@noop {} {\bibfield  {journal}
  {\bibinfo  {journal} {Journal of Physics: Condensed Matter}\ }\textbf
  {\bibinfo {volume} {32}},\ \bibinfo {pages} {015901} (\bibinfo {year}
  {2020})}\BibitemShut {NoStop}%
\bibitem [{\citenamefont {Geng}\ \emph {et~al.}(2025)\citenamefont {Geng},
  \citenamefont {Liu}, \citenamefont {Xu} \emph {et~al.}}]{Geng2025VASPKIT}%
  \BibitemOpen
  \bibfield  {author} {\bibinfo {author} {\bibfnamefont {W.~T.}\ \bibnamefont
  {Geng}}, \bibinfo {author} {\bibfnamefont {Y.~C.}\ \bibnamefont {Liu}},
  \bibinfo {author} {\bibfnamefont {N.}~\bibnamefont {Xu}}, \emph {et~al.},\
  }\bibfield  {title} {\bibinfo {title} {Empowering materials science with
  vaspkit: a toolkit for enhanced simulation and analysis},\ }\bibfield
  {journal} {\bibinfo  {journal} {Nature Protocols}\ }\href
  {https://doi.org/10.1038/s41596-025-01160-w} {10.1038/s41596-025-01160-w}
  (\bibinfo {year} {2025})\BibitemShut {NoStop}%
\bibitem [{\citenamefont {Wang}\ \emph {et~al.}(2021)\citenamefont {Wang},
  \citenamefont {Xu}, \citenamefont {Liu}, \citenamefont {Tang},\ and\
  \citenamefont {Geng}}]{Wang2021VASPKIT}%
  \BibitemOpen
  \bibfield  {author} {\bibinfo {author} {\bibfnamefont {V.}~\bibnamefont
  {Wang}}, \bibinfo {author} {\bibfnamefont {N.}~\bibnamefont {Xu}}, \bibinfo
  {author} {\bibfnamefont {J.~C.}\ \bibnamefont {Liu}}, \bibinfo {author}
  {\bibfnamefont {G.}~\bibnamefont {Tang}},\ and\ \bibinfo {author}
  {\bibfnamefont {W.~T.}\ \bibnamefont {Geng}},\ }\bibfield  {title} {\bibinfo
  {title} {Vaspkit: A user-friendly interface facilitating high-throughput
  computing and analysis using vasp code},\ }\href
  {https://doi.org/10.1016/j.cpc.2021.108033} {\bibfield  {journal} {\bibinfo
  {journal} {Computer Physics Communications}\ }\textbf {\bibinfo {volume}
  {267}},\ \bibinfo {pages} {108033} (\bibinfo {year} {2021})}\BibitemShut
  {NoStop}%
\bibitem [{\citenamefont {Hong}\ and\ \citenamefont {van~de
  Walle}(2013)}]{Hong2013SmallCoex}%
  \BibitemOpen
  \bibfield  {author} {\bibinfo {author} {\bibfnamefont {Q.-J.}\ \bibnamefont
  {Hong}}\ and\ \bibinfo {author} {\bibfnamefont {A.}~\bibnamefont {van~de
  Walle}},\ }\bibfield  {title} {\bibinfo {title} {Solid--liquid coexistence in
  small systems: A statistical method to calculate melting temperatures},\
  }\href {https://doi.org/10.1063/1.4819896} {\bibfield  {journal} {\bibinfo
  {journal} {The Journal of Chemical Physics}\ }\textbf {\bibinfo {volume}
  {139}},\ \bibinfo {pages} {094114} (\bibinfo {year} {2013})}\BibitemShut
  {NoStop}%
\bibitem [{\citenamefont {Hong}\ and\ \citenamefont {van~de
  Walle}(2016)}]{Hong2016}%
  \BibitemOpen
  \bibfield  {author} {\bibinfo {author} {\bibfnamefont {Q.-J.}\ \bibnamefont
  {Hong}}\ and\ \bibinfo {author} {\bibfnamefont {A.}~\bibnamefont {van~de
  Walle}},\ }\bibfield  {title} {\bibinfo {title} {A user guide for sluschi:
  Solid and liquid in ultra small coexistence with hovering interfaces},\
  }\href {https://doi.org/10.1016/j.calphad.2015.12.003} {\bibfield  {journal}
  {\bibinfo  {journal} {CALPHAD: Computer Coupling of Phase Diagrams and
  Thermochemistry}\ }\textbf {\bibinfo {volume} {52}},\ \bibinfo {pages} {88}
  (\bibinfo {year} {2016})}\BibitemShut {NoStop}%
\bibitem [{\citenamefont {Flyvbjerg}\ and\ \citenamefont
  {Petersen}(1989)}]{Flyvbjerg1989BlockAverage}%
  \BibitemOpen
  \bibfield  {author} {\bibinfo {author} {\bibfnamefont {H.}~\bibnamefont
  {Flyvbjerg}}\ and\ \bibinfo {author} {\bibfnamefont {H.~G.}\ \bibnamefont
  {Petersen}},\ }\bibfield  {title} {\bibinfo {title} {Error estimates on
  averages of correlated data},\ }\href {https://doi.org/10.1063/1.457480}
  {\bibfield  {journal} {\bibinfo  {journal} {The Journal of Chemical Physics}\
  }\textbf {\bibinfo {volume} {91}},\ \bibinfo {pages} {461} (\bibinfo {year}
  {1989})}\BibitemShut {NoStop}%
\bibitem [{\citenamefont {Frenkel}\ and\ \citenamefont
  {Smit}(2002)}]{FrenkelSmit2002Book}%
  \BibitemOpen
  \bibfield  {author} {\bibinfo {author} {\bibfnamefont {D.}~\bibnamefont
  {Frenkel}}\ and\ \bibinfo {author} {\bibfnamefont {B.}~\bibnamefont {Smit}},\
  }\href@noop {} {\emph {\bibinfo {title} {Understanding Molecular Simulation:
  From Algorithms to Applications}}},\ \bibinfo {edition} {2nd}\ ed.,\
  Computational Science Series\ (\bibinfo  {publisher} {Academic Press},\
  \bibinfo {address} {San Diego, California},\ \bibinfo {year}
  {2002})\BibitemShut {NoStop}%
\bibitem [{\citenamefont {Schick}\ \emph {et~al.}(2012)\citenamefont {Schick},
  \citenamefont {Brillo}, \citenamefont {Egry},\ and\ \citenamefont
  {Hallstedt}}]{Schick2012}%
  \BibitemOpen
  \bibfield  {author} {\bibinfo {author} {\bibfnamefont {M.}~\bibnamefont
  {Schick}}, \bibinfo {author} {\bibfnamefont {J.}~\bibnamefont {Brillo}},
  \bibinfo {author} {\bibfnamefont {I.}~\bibnamefont {Egry}},\ and\ \bibinfo
  {author} {\bibfnamefont {B.}~\bibnamefont {Hallstedt}},\ }\bibfield  {title}
  {\bibinfo {title} {Viscosity of al--cu liquid alloys: measurement and
  thermodynamic description},\ }\href
  {https://doi.org/10.1007/s10853-012-6710-x} {\bibfield  {journal} {\bibinfo
  {journal} {Journal of Materials Science}\ }\textbf {\bibinfo {volume} {47}},\
  \bibinfo {pages} {7938} (\bibinfo {year} {2012})}\BibitemShut {NoStop}%
\bibitem [{\citenamefont {{Thermo-Calc Software AB}}(2025)}]{TCAL9_2025}%
  \BibitemOpen
  \bibfield  {author} {\bibinfo {author} {\bibnamefont {{Thermo-Calc Software
  AB}}},\ }\href@noop {} {\bibinfo {title} {{TCAL9} -- thermo-calc
  aluminum-based alloys database}},\ \bibinfo {howpublished}
  {\url{https://thermocalc.com/products/databases/aluminum-based-alloys/}}
  (\bibinfo {year} {2025}),\ \bibinfo {note} {accessed
  16~October~2025}\BibitemShut {NoStop}%
\bibitem [{\citenamefont {Costigliola}\ \emph {et~al.}(2019)\citenamefont
  {Costigliola}, \citenamefont {Heyes}, \citenamefont {Schr{\o}der},\ and\
  \citenamefont {Dyre}}]{Costigliola2019}%
  \BibitemOpen
  \bibfield  {author} {\bibinfo {author} {\bibfnamefont {L.}~\bibnamefont
  {Costigliola}}, \bibinfo {author} {\bibfnamefont {D.~M.}\ \bibnamefont
  {Heyes}}, \bibinfo {author} {\bibfnamefont {T.~B.}\ \bibnamefont
  {Schr{\o}der}},\ and\ \bibinfo {author} {\bibfnamefont {J.~C.}\ \bibnamefont
  {Dyre}},\ }\bibfield  {title} {\bibinfo {title} {Revisiting the
  stokes--einstein relation without a hydrodynamic diameter},\ }\href
  {https://doi.org/10.1063/1.5085615} {\bibfield  {journal} {\bibinfo
  {journal} {The Journal of Chemical Physics}\ }\textbf {\bibinfo {volume}
  {150}},\ \bibinfo {pages} {021101} (\bibinfo {year} {2019})}\BibitemShut
  {NoStop}%
\bibitem [{\citenamefont {Pan}\ \emph {et~al.}(2017)\citenamefont {Pan},
  \citenamefont {Wang}, \citenamefont {Wang},\ and\ \citenamefont
  {Bai}}]{Pan2017}%
  \BibitemOpen
  \bibfield  {author} {\bibinfo {author} {\bibfnamefont {S.}~\bibnamefont
  {Pan}}, \bibinfo {author} {\bibfnamefont {Z.}~\bibnamefont {Wang}}, \bibinfo
  {author} {\bibfnamefont {W.-H.}\ \bibnamefont {Wang}},\ and\ \bibinfo
  {author} {\bibfnamefont {H.}~\bibnamefont {Bai}},\ }\bibfield  {title}
  {\bibinfo {title} {A structural signature of the breakdown of the
  stokes--einstein relation in metallic liquids},\ }\href
  {https://doi.org/10.1039/C7CP03475J} {\bibfield  {journal} {\bibinfo
  {journal} {Physical Chemistry Chemical Physics}\ }\textbf {\bibinfo {volume}
  {19}},\ \bibinfo {pages} {26271} (\bibinfo {year} {2017})}\BibitemShut
  {NoStop}%
\bibitem [{\citenamefont {Singh}\ \emph {et~al.}(2021)\citenamefont {Singh},
  \citenamefont {Tiwari}, \citenamefont {Sharma},\ and\ \citenamefont
  {Srivastava}}]{FeCoNi2021}%
  \BibitemOpen
  \bibfield  {author} {\bibinfo {author} {\bibfnamefont {S.}~\bibnamefont
  {Singh}}, \bibinfo {author} {\bibfnamefont {R.~S.}\ \bibnamefont {Tiwari}},
  \bibinfo {author} {\bibfnamefont {B.~K.}\ \bibnamefont {Sharma}},\ and\
  \bibinfo {author} {\bibfnamefont {S.}~\bibnamefont {Srivastava}},\ }\bibfield
   {title} {\bibinfo {title} {Validity of the stokes--einstein relation in
  liquid 3d transition metals (fe, co, ni)},\ }\href
  {https://doi.org/10.1016/j.physa.2021.125969} {\bibfield  {journal} {\bibinfo
   {journal} {Physica A: Statistical Mechanics and its Applications}\ }\textbf
  {\bibinfo {volume} {574}},\ \bibinfo {pages} {125969} (\bibinfo {year}
  {2021})}\BibitemShut {NoStop}%
\bibitem [{\citenamefont {Perdew}\ \emph {et~al.}(1996)\citenamefont {Perdew},
  \citenamefont {Burke},\ and\ \citenamefont {Ernzerhof}}]{Perdew1996PBE}%
  \BibitemOpen
  \bibfield  {author} {\bibinfo {author} {\bibfnamefont {J.~P.}\ \bibnamefont
  {Perdew}}, \bibinfo {author} {\bibfnamefont {K.}~\bibnamefont {Burke}},\ and\
  \bibinfo {author} {\bibfnamefont {M.}~\bibnamefont {Ernzerhof}},\ }\bibfield
  {title} {\bibinfo {title} {Generalized gradient approximation made simple},\
  }\href {https://doi.org/10.1103/PhysRevLett.77.3865} {\bibfield  {journal}
  {\bibinfo  {journal} {Physical Review Letters}\ }\textbf {\bibinfo {volume}
  {77}},\ \bibinfo {pages} {3865} (\bibinfo {year} {1996})}\BibitemShut
  {NoStop}%
\bibitem [{\citenamefont {Geiger}\ \emph {et~al.}(2011)\citenamefont {Geiger},
  \citenamefont {Alekseev}, \citenamefont {Lazic}, \citenamefont {Fisch},
  \citenamefont {Armbruster}, \citenamefont {Langner}, \citenamefont
  {Fechtelkord}, \citenamefont {Kim}, \citenamefont {Pettke},\ and\
  \citenamefont {Weppner}}]{Geiger2011}%
  \BibitemOpen
  \bibfield  {author} {\bibinfo {author} {\bibfnamefont {C.~A.}\ \bibnamefont
  {Geiger}}, \bibinfo {author} {\bibfnamefont {E.~V.}\ \bibnamefont
  {Alekseev}}, \bibinfo {author} {\bibfnamefont {B.}~\bibnamefont {Lazic}},
  \bibinfo {author} {\bibfnamefont {M.}~\bibnamefont {Fisch}}, \bibinfo
  {author} {\bibfnamefont {T.}~\bibnamefont {Armbruster}}, \bibinfo {author}
  {\bibfnamefont {R.}~\bibnamefont {Langner}}, \bibinfo {author} {\bibfnamefont
  {M.}~\bibnamefont {Fechtelkord}}, \bibinfo {author} {\bibfnamefont
  {N.}~\bibnamefont {Kim}}, \bibinfo {author} {\bibfnamefont {T.}~\bibnamefont
  {Pettke}},\ and\ \bibinfo {author} {\bibfnamefont {W.}~\bibnamefont
  {Weppner}},\ }\bibfield  {title} {\bibinfo {title} {Crystal chemistry and
  stability of li7la3zr2o12 garnet: A fast lithium-ion conductor},\ }\href
  {https://doi.org/10.1021/ic101914e} {\bibfield  {journal} {\bibinfo
  {journal} {Inorganic Chemistry}\ }\textbf {\bibinfo {volume} {50}},\ \bibinfo
  {pages} {1089} (\bibinfo {year} {2011})}\BibitemShut {NoStop}%
\bibitem [{\citenamefont {Matsui}\ \emph {et~al.}(2014)\citenamefont {Matsui},
  \citenamefont {Sakamoto}, \citenamefont {Takahashi}, \citenamefont {Hirano},
  \citenamefont {Takeda}, \citenamefont {Yamamoto},\ and\ \citenamefont
  {Imanishi}}]{Matsui2014}%
  \BibitemOpen
  \bibfield  {author} {\bibinfo {author} {\bibfnamefont {M.}~\bibnamefont
  {Matsui}}, \bibinfo {author} {\bibfnamefont {K.}~\bibnamefont {Sakamoto}},
  \bibinfo {author} {\bibfnamefont {K.}~\bibnamefont {Takahashi}}, \bibinfo
  {author} {\bibfnamefont {A.}~\bibnamefont {Hirano}}, \bibinfo {author}
  {\bibfnamefont {Y.}~\bibnamefont {Takeda}}, \bibinfo {author} {\bibfnamefont
  {O.}~\bibnamefont {Yamamoto}},\ and\ \bibinfo {author} {\bibfnamefont
  {N.}~\bibnamefont {Imanishi}},\ }\bibfield  {title} {\bibinfo {title} {Phase
  transformation of the garnet structured lithium ion conductor:
  Li7la3zr2o12},\ }\href {https://doi.org/10.1016/j.ssi.2013.09.027} {\bibfield
   {journal} {\bibinfo  {journal} {Solid State Ionics}\ }\textbf {\bibinfo
  {volume} {262}},\ \bibinfo {pages} {155} (\bibinfo {year}
  {2014})}\BibitemShut {NoStop}%
\bibitem [{\citenamefont {Hong}\ \emph {et~al.}(2024)\citenamefont {Hong},
  \citenamefont {Tepesch},\ and\ \citenamefont {van~de
  Walle}}]{Hong2024LLZO_PhaseDiagram}%
  \BibitemOpen
  \bibfield  {author} {\bibinfo {author} {\bibfnamefont {Q.-J.}\ \bibnamefont
  {Hong}}, \bibinfo {author} {\bibfnamefont {P.~D.}\ \bibnamefont {Tepesch}},\
  and\ \bibinfo {author} {\bibfnamefont {A.}~\bibnamefont {van~de Walle}},\
  }\bibfield  {title} {\bibinfo {title} {Combined experimental and
  computational assessment of the {Li$_2$O--La$_2$O$_3$--ZrO$_2$} phase
  diagram},\ }\href {https://doi.org/10.1111/jace.19568} {\bibfield  {journal}
  {\bibinfo  {journal} {Journal of the American Ceramic Society}\ }\textbf
  {\bibinfo {volume} {107}},\ \bibinfo {pages} {5682} (\bibinfo {year}
  {2024})}\BibitemShut {NoStop}%
\bibitem [{\citenamefont {Bernstein}\ \emph {et~al.}(2012)\citenamefont
  {Bernstein}, \citenamefont {Johannes},\ and\ \citenamefont
  {Hoang}}]{Bernstein2012}%
  \BibitemOpen
  \bibfield  {author} {\bibinfo {author} {\bibfnamefont {N.}~\bibnamefont
  {Bernstein}}, \bibinfo {author} {\bibfnamefont {M.~D.}\ \bibnamefont
  {Johannes}},\ and\ \bibinfo {author} {\bibfnamefont {K.}~\bibnamefont
  {Hoang}},\ }\bibfield  {title} {\bibinfo {title} {Origin of the structural
  phase transition in li7la3zr2o12},\ }\href
  {https://doi.org/10.1103/PhysRevLett.109.205702} {\bibfield  {journal}
  {\bibinfo  {journal} {Physical Review Letters}\ }\textbf {\bibinfo {volume}
  {109}},\ \bibinfo {pages} {205702} (\bibinfo {year} {2012})}\BibitemShut
  {NoStop}%
\bibitem [{\citenamefont {Meier}\ \emph {et~al.}(2014)\citenamefont {Meier},
  \citenamefont {Laino},\ and\ \citenamefont {Curioni}}]{Meier2014}%
  \BibitemOpen
  \bibfield  {author} {\bibinfo {author} {\bibfnamefont {K.}~\bibnamefont
  {Meier}}, \bibinfo {author} {\bibfnamefont {T.}~\bibnamefont {Laino}},\ and\
  \bibinfo {author} {\bibfnamefont {A.}~\bibnamefont {Curioni}},\ }\bibfield
  {title} {\bibinfo {title} {Solid-state electrolytes: Revealing the mechanisms
  of li-ion conduction in tetragonal and cubic llzo by first-principles
  calculations},\ }\href {https://doi.org/10.1021/jp5002463} {\bibfield
  {journal} {\bibinfo  {journal} {Journal of Physical Chemistry C}\ }\textbf
  {\bibinfo {volume} {118}},\ \bibinfo {pages} {6668} (\bibinfo {year}
  {2014})}\BibitemShut {NoStop}%
\bibitem [{\citenamefont {Klenk}\ and\ \citenamefont {Lai}(2015)}]{Klenk2015}%
  \BibitemOpen
  \bibfield  {author} {\bibinfo {author} {\bibfnamefont {M.~J.}\ \bibnamefont
  {Klenk}}\ and\ \bibinfo {author} {\bibfnamefont {W.}~\bibnamefont {Lai}},\
  }\bibfield  {title} {\bibinfo {title} {Local structure and dynamics of
  lithium garnet ionic conductors: Tetragonal and cubic li7la3zr2o12},\ }\href
  {https://doi.org/10.1039/C4CP05690F} {\bibfield  {journal} {\bibinfo
  {journal} {Physical Chemistry Chemical Physics}\ }\textbf {\bibinfo {volume}
  {17}},\ \bibinfo {pages} {8758} (\bibinfo {year} {2015})}\BibitemShut
  {NoStop}%
\bibitem [{\citenamefont {Tian}\ \emph {et~al.}(2023)\citenamefont {Tian},
  \citenamefont {Cai}, \citenamefont {Tan}, \citenamefont {Lin}, \citenamefont
  {Phillips}, \citenamefont {Abrahams}, \citenamefont {Keen}, \citenamefont
  {Keeble}, \citenamefont {Fiedler}, \citenamefont {Zhang}, \citenamefont
  {Kong},\ and\ \citenamefont {Dove}}]{Tian2023}%
  \BibitemOpen
  \bibfield  {author} {\bibinfo {author} {\bibfnamefont {H.}~\bibnamefont
  {Tian}}, \bibinfo {author} {\bibfnamefont {G.}~\bibnamefont {Cai}}, \bibinfo
  {author} {\bibfnamefont {L.}~\bibnamefont {Tan}}, \bibinfo {author}
  {\bibfnamefont {H.}~\bibnamefont {Lin}}, \bibinfo {author} {\bibfnamefont
  {A.~E.}\ \bibnamefont {Phillips}}, \bibinfo {author} {\bibfnamefont
  {I.}~\bibnamefont {Abrahams}}, \bibinfo {author} {\bibfnamefont {D.~A.}\
  \bibnamefont {Keen}}, \bibinfo {author} {\bibfnamefont {D.~S.}\ \bibnamefont
  {Keeble}}, \bibinfo {author} {\bibfnamefont {A.}~\bibnamefont {Fiedler}},
  \bibinfo {author} {\bibfnamefont {J.}~\bibnamefont {Zhang}}, \bibinfo
  {author} {\bibfnamefont {X.~Y.}\ \bibnamefont {Kong}},\ and\ \bibinfo
  {author} {\bibfnamefont {M.~T.}\ \bibnamefont {Dove}},\ }\bibfield  {title}
  {\bibinfo {title} {Local structure and lithium-ion diffusion pathway of cubic
  li7la3zr2o12 studied by total scattering and the reverse monte carlo
  method},\ }\href {https://doi.org/10.1039/D3TA04495E} {\bibfield  {journal}
  {\bibinfo  {journal} {Journal of Materials Chemistry A}\ }\textbf {\bibinfo
  {volume} {11}},\ \bibinfo {pages} {25516} (\bibinfo {year}
  {2023})}\BibitemShut {NoStop}%
\bibitem [{\citenamefont {Wang}\ \emph {et~al.}(2024)\citenamefont {Wang},
  \citenamefont {Ushakov}, \citenamefont {Opila}, \citenamefont {Navrotsky},\
  and\ \citenamefont {Hong}}]{Wang2024Acta}%
  \BibitemOpen
  \bibfield  {author} {\bibinfo {author} {\bibfnamefont {L.}~\bibnamefont
  {Wang}}, \bibinfo {author} {\bibfnamefont {S.~V.}\ \bibnamefont {Ushakov}},
  \bibinfo {author} {\bibfnamefont {E.~J.}\ \bibnamefont {Opila}}, \bibinfo
  {author} {\bibfnamefont {A.}~\bibnamefont {Navrotsky}},\ and\ \bibinfo
  {author} {\bibfnamefont {Q.-J.}\ \bibnamefont {Hong}},\ }\bibfield  {title}
  {\bibinfo {title} {High-temperature crystal structure prediction from
  \textit{ab initio} molecular dynamics with sluschi},\ }\href
  {https://doi.org/10.1016/j.actamat.2024.120432} {\bibfield  {journal}
  {\bibinfo  {journal} {Acta Materialia}\ }\textbf {\bibinfo {volume} {281}},\
  \bibinfo {pages} {120432} (\bibinfo {year} {2024})}\BibitemShut {NoStop}%
\bibitem [{\citenamefont {Ushakov}\ \emph {et~al.}(2024)\citenamefont
  {Ushakov}, \citenamefont {Hong}, \citenamefont {III}, \citenamefont {van~de
  Walle},\ and\ \citenamefont {Navrotsky}}]{Ushakov2024Er2O3_Tm2O3}%
  \BibitemOpen
  \bibfield  {author} {\bibinfo {author} {\bibfnamefont {S.~V.}\ \bibnamefont
  {Ushakov}}, \bibinfo {author} {\bibfnamefont {Q.-J.}\ \bibnamefont {Hong}},
  \bibinfo {author} {\bibfnamefont {A.~P.}\ \bibnamefont {III}}, \bibinfo
  {author} {\bibfnamefont {A.}~\bibnamefont {van~de Walle}},\ and\ \bibinfo
  {author} {\bibfnamefont {A.}~\bibnamefont {Navrotsky}},\ }\bibfield  {title}
  {\bibinfo {title} {Thermal expansion and enthalpies of phase transformation
  and fusion of {Er$_2$O$_3$} and {Tm$_2$O$_3$} from experiment and
  computation},\ }\href {https://doi.org/10.1021/acs.chemmater.4c00477}
  {\bibfield  {journal} {\bibinfo  {journal} {Chemistry of Materials}\ }\textbf
  {\bibinfo {volume} {36}},\ \bibinfo {pages} {4868} (\bibinfo {year}
  {2024})}\BibitemShut {NoStop}%
\bibitem [{\citenamefont {Hong}\ and\ \citenamefont {van~de
  Walle}(2015)}]{Hong2015HfC_TaC_Melting}%
  \BibitemOpen
  \bibfield  {author} {\bibinfo {author} {\bibfnamefont {Q.-J.}\ \bibnamefont
  {Hong}}\ and\ \bibinfo {author} {\bibfnamefont {A.}~\bibnamefont {van~de
  Walle}},\ }\bibfield  {title} {\bibinfo {title} {Prediction of the material
  with highest known melting point from \textit{ab initio} molecular dynamics
  calculations},\ }\href {https://doi.org/10.1103/PhysRevB.92.020104}
  {\bibfield  {journal} {\bibinfo  {journal} {Physical Review B}\ }\textbf
  {\bibinfo {volume} {92}},\ \bibinfo {pages} {020104} (\bibinfo {year}
  {2015})}\BibitemShut {NoStop}%
\bibitem [{\citenamefont {Hong}\ and\ \citenamefont
  {Liu}(2025)}]{HongLiu2025Entropy}%
  \BibitemOpen
  \bibfield  {author} {\bibinfo {author} {\bibfnamefont {Q.-J.}\ \bibnamefont
  {Hong}}\ and\ \bibinfo {author} {\bibfnamefont {Z.-K.}\ \bibnamefont {Liu}},\
  }\bibfield  {title} {\bibinfo {title} {Generalized approach for rapid entropy
  calculation of liquids and solids},\ }\href
  {https://doi.org/10.1103/PhysRevResearch.7.L012030} {\bibfield  {journal}
  {\bibinfo  {journal} {Physical Review Research}\ }\textbf {\bibinfo {volume}
  {7}},\ \bibinfo {pages} {L012030} (\bibinfo {year} {2025})}\BibitemShut
  {NoStop}%
\bibitem [{\citenamefont {Takada}\ and\ \citenamefont
  {Adachi}(1986)}]{Takada1986-Alpha}%
  \BibitemOpen
  \bibfield  {author} {\bibinfo {author} {\bibfnamefont {J.}~\bibnamefont
  {Takada}}\ and\ \bibinfo {author} {\bibfnamefont {M.}~\bibnamefont
  {Adachi}},\ }\bibfield  {title} {\bibinfo {title} {Determination of diffusion
  coefficient of oxygen in alpha-iron from internal oxidation measurements in
  fe–si alloys},\ }\href {https://doi.org/10.1007/BF00547959} {\bibfield
  {journal} {\bibinfo  {journal} {Journal of Materials Science}\ }\textbf
  {\bibinfo {volume} {21}},\ \bibinfo {pages} {2133} (\bibinfo {year}
  {1986})}\BibitemShut {NoStop}%
\bibitem [{\citenamefont {Takada}\ \emph {et~al.}(1986)\citenamefont {Takada},
  \citenamefont {Yamamoto}, \citenamefont {Kikuchi},\ and\ \citenamefont
  {Adachi}}]{Takada1986-Gamma}%
  \BibitemOpen
  \bibfield  {author} {\bibinfo {author} {\bibfnamefont {J.}~\bibnamefont
  {Takada}}, \bibinfo {author} {\bibfnamefont {S.}~\bibnamefont {Yamamoto}},
  \bibinfo {author} {\bibfnamefont {S.}~\bibnamefont {Kikuchi}},\ and\ \bibinfo
  {author} {\bibfnamefont {M.}~\bibnamefont {Adachi}},\ }\bibfield  {title}
  {\bibinfo {title} {Determination of diffusion coefficient of oxygen in
  gamma-iron from measurements of internal oxidation in fe–al alloys},\
  }\href@noop {} {\bibfield  {journal} {\bibinfo  {journal} {Metallurgical
  Transactions A}\ }\textbf {\bibinfo {volume} {17}},\ \bibinfo {pages} {221}
  (\bibinfo {year} {1986})}\BibitemShut {NoStop}%
\bibitem [{\citenamefont {Mohanta}\ \emph
  {et~al.}(2025{\natexlab{a}})\citenamefont {Mohanta}, \citenamefont {Fisher},
  \citenamefont {Korobeinikov}, \citenamefont {Hong}, \citenamefont {Muhich},
  \citenamefont {Seetharaman},\ and\ \citenamefont
  {Leick}}]{Mohanta2025HydrogenReduction}%
  \BibitemOpen
  \bibfield  {author} {\bibinfo {author} {\bibfnamefont {R.~K.}\ \bibnamefont
  {Mohanta}}, \bibinfo {author} {\bibfnamefont {D.}~\bibnamefont {Fisher}},
  \bibinfo {author} {\bibfnamefont {Y.}~\bibnamefont {Korobeinikov}}, \bibinfo
  {author} {\bibfnamefont {Q.-J.}\ \bibnamefont {Hong}}, \bibinfo {author}
  {\bibfnamefont {C.}~\bibnamefont {Muhich}}, \bibinfo {author} {\bibfnamefont
  {S.}~\bibnamefont {Seetharaman}},\ and\ \bibinfo {author} {\bibfnamefont
  {N.}~\bibnamefont {Leick}},\ }\bibfield  {title} {\bibinfo {title} {Bridging
  the gap in carbon-free iron making: How hydrogen affects the reduction of
  iron ore between 900 and 1590~\textdegree c},\ }\href
  {https://doi.org/10.1021/acssuschemeng.5c04984} {\bibfield  {journal}
  {\bibinfo  {journal} {ACS Sustainable Chemistry \& Engineering}\ }\textbf
  {\bibinfo {volume} {13}},\ \bibinfo {pages} {16237} (\bibinfo {year}
  {2025}{\natexlab{a}})},\ \bibinfo {note} {open Access, Letter}\BibitemShut
  {NoStop}%
\bibitem [{\citenamefont {Dunn}(1983)}]{Dunn1983}%
  \BibitemOpen
  \bibfield  {author} {\bibinfo {author} {\bibfnamefont {T.}~\bibnamefont
  {Dunn}},\ }\bibfield  {title} {\bibinfo {title} {Oxygen chemical diffusion in
  three basaltic liquids at elevated temperatures and pressures},\ }\href
  {https://doi.org/10.1016/0016-7037(83)90205-9} {\bibfield  {journal}
  {\bibinfo  {journal} {Geochimica et Cosmochimica Acta}\ }\textbf {\bibinfo
  {volume} {47}},\ \bibinfo {pages} {1923} (\bibinfo {year}
  {1983})}\BibitemShut {NoStop}%
\bibitem [{\citenamefont {Mohanta}\ \emph
  {et~al.}(2025{\natexlab{b}})\citenamefont {Mohanta}, \citenamefont {Fisher},
  \citenamefont {Korobeinikov}, \citenamefont {Hong}, \citenamefont {Leick},\
  and\ \citenamefont {Seetharaman}}]{Mohanta2025MeltThenReduce}%
  \BibitemOpen
  \bibfield  {author} {\bibinfo {author} {\bibfnamefont {R.~K.}\ \bibnamefont
  {Mohanta}}, \bibinfo {author} {\bibfnamefont {D.}~\bibnamefont {Fisher}},
  \bibinfo {author} {\bibfnamefont {Y.}~\bibnamefont {Korobeinikov}}, \bibinfo
  {author} {\bibfnamefont {Q.-J.}\ \bibnamefont {Hong}}, \bibinfo {author}
  {\bibfnamefont {N.}~\bibnamefont {Leick}},\ and\ \bibinfo {author}
  {\bibfnamefont {S.}~\bibnamefont {Seetharaman}},\ }\href
  {https://doi.org/10.26434/chemrxiv-2025-chqng} {\bibinfo {title} {Hydrogen
  driven reduction of dri-grade iron ore above 1500~\textdegree c: a
  ``melt-then-reduce'' approach}},\ \bibinfo {howpublished} {ChemRxiv Preprint}
  (\bibinfo {year} {2025}{\natexlab{b}}),\ \bibinfo {note} {version~1,
  17~July~2025}\BibitemShut {NoStop}%
\bibitem [{\citenamefont {Dunn}\ and\ \citenamefont
  {Scarfe}(1986)}]{Dunn1986-ChemGeol}%
  \BibitemOpen
  \bibfield  {author} {\bibinfo {author} {\bibfnamefont {T.}~\bibnamefont
  {Dunn}}\ and\ \bibinfo {author} {\bibfnamefont {C.~M.}\ \bibnamefont
  {Scarfe}},\ }\bibfield  {title} {\bibinfo {title} {Variation of the chemical
  diffusivity of oxygen and viscosity of an andesite melt with pressure at
  constant temperature},\ }\href {https://doi.org/10.1016/0009-2541(86)90119-4}
  {\bibfield  {journal} {\bibinfo  {journal} {Chemical Geology}\ }\textbf
  {\bibinfo {volume} {54}},\ \bibinfo {pages} {203} (\bibinfo {year}
  {1986})}\BibitemShut {NoStop}%
\bibitem [{\citenamefont {Pozzo}\ \emph {et~al.}(2013)\citenamefont {Pozzo},
  \citenamefont {Davies}, \citenamefont {Gubbins},\ and\ \citenamefont
  {Alf{\`e}}}]{Pozzo2013}%
  \BibitemOpen
  \bibfield  {author} {\bibinfo {author} {\bibfnamefont {M.}~\bibnamefont
  {Pozzo}}, \bibinfo {author} {\bibfnamefont {C.}~\bibnamefont {Davies}},
  \bibinfo {author} {\bibfnamefont {D.}~\bibnamefont {Gubbins}},\ and\ \bibinfo
  {author} {\bibfnamefont {D.}~\bibnamefont {Alf{\`e}}},\ }\bibfield  {title}
  {\bibinfo {title} {Transport properties for liquid silicon--oxygen--iron
  mixtures at earth’s core conditions},\ }\href
  {https://doi.org/10.1103/PhysRevB.87.014110} {\bibfield  {journal} {\bibinfo
  {journal} {Physical Review B}\ }\textbf {\bibinfo {volume} {87}},\ \bibinfo
  {pages} {014110} (\bibinfo {year} {2013})}\BibitemShut {NoStop}%
\end{thebibliography}%

\end{document}